\documentclass[traditabstract]{aa} 
\usepackage{array}
\usepackage{amssymb}
\usepackage{graphicx}
\usepackage{natbib}
\usepackage{lscape}
\usepackage{rotating}
\usepackage{longtable}
\newcommand{\xmm}{{\sc{XMM}}\emph{-Newton}}
\newcommand{\swift}{\em{Swift}}
\begin{document}
\title{The X-ray light curve of the massive colliding wind \\
Wolf-Rayet + O binary WR\,21a\thanks{Based on observations collected at ESO as well as with {\it{Swift}}, {\it{Chandra}}, and the ESA science mission \xmm , an ESA Science Mission with instruments and contributions directly funded by ESA Member States and the USA (NASA).}}

\author{Eric~Gosset\thanks{F.R.S.-FNRS Senior Research Associate.}
\and Ya\"el~Naz\'e\thanks{F.R.S.-FNRS Research Associate.}
}

\institute{Groupe d'Astrophysique des Hautes Energies, Institut d'Astrophysique 
et de G\'eophysique, Universit\'e de Li\`ege, Quartier Agora (B5c), 
All\'ee du 6 Ao\^ut 19c, B-4000 Sart Tilman, Li\`ege, Belgium\\
}

\authorrunning{Gosset \& Naz\'e}
\titlerunning{X-ray light curve of WR\,21a}

\abstract{Our dedicated \xmm\ monitoring, as well as archival 
{\it{Chandra}} and {\it{Swift}} datasets, were used to examine 
the behaviour of the WN5h+O3V binary WR\,21a at high energies. 
For most of the orbit, the X-ray emission exhibits few variations. 
However, an increase in strength of the emission is seen before 
periastron, following a $1/D$ relative trend, where $D$ is the 
separation between both components. This increase is rapidly 
followed by a decline due to strong absorption as the 
Wolf-Rayet (WR) comes in front. The fitted local absorption value 
appears to be coherent with a mass-loss rate of about 1$\times$10$^{-5}$ M$_{\sun}$ yr$^{-1}$ for the WR component. 
However, absorption is not the only parameter affecting the 
X-ray emission at periastron as even the hard X-ray 
emission  decreases, suggesting a possible collapse of the 
colliding wind region near to or onto the photosphere of the 
companion just before or at periastron. An eclipse may appear 
as another potential scenario, but it would be in apparent 
contradiction with several lines of evidence, notably the width 
of the dip in the X-ray light curve and the absence of variations 
in the UV light curve. Afterwards, the emission slowly recovers, 
with a strong hysteresis effect. The observed behaviour is 
compatible with predictions from general wind-wind collision models 
although the absorption increase is too shallow.}
\maketitle
\section{Introduction}
Massive stars of type O and early B, and 
Wolf-Rayet (WR) stars, their evolved descendants,
are very important objects that have a large 
impact on their host galaxy. Indeed, they participate in the chemical 
evolution of the interstellar medium and dominate the mechanical 
evolution of their surroundings by carving bubbles 
and influencing star formation. 
Despite this importance, the main fundamental physical
parameters characterising them remain poorly known and the actual 
details of massive star evolution are yet to be understood 
(e.g.\ the luminous blue variable phase, the effect of rotation and
the corresponding internal law).
This is particularly true for the most massive objects.

Certainly the main basic parameter is the mass.
Classically, astronomers supposed that the most 
massive stars were to be found amongst very early (O2--O3) stars. 
However, recent clues tend to prove that the most massive stars evolve 
rapidly towards core-hydrogen-burning objects 
appearing as disguised hydrogen-rich WR stars, 
of the WNLh (late H-rich WN) type.
In that context, binary system investigations played a key role in 
recognising the true nature of WNLh stars. 
The relatively small number of extremely massive Galactic stars 
implies that the discovery and in-depth study of any such system provide 
breakthrough information bringing new constraints on stellar evolution.

The second crucial parameter for massive star evolution is the 
mass-loss rate, which remains not well known, mainly due to uncertain 
clumping properties. Therefore, a large palette of possible values 
are usually obtained for any star. In a massive binary system, 
the winds of both stars collide in a so-called colliding wind region (CWR) 
broadly located between the two stars. In some cases, a plasma at high 
temperature (some 10$^7$\,K) is generated, which then emits an intense 
thermal X-ray emission in addition to the intrinsic 
emission\footnote{Since the 
discovery by the {\it{EINSTEIN}} satellite that massive 
stars could be moderate X-ray emitters \citep{harn79, sewa79}, 
it appears that the observed X-ray luminosity of single OB stars is 
proportional to their bolometric luminosity with an observed 
ratio around 10$^{-7}$
\citep{pall81, sech82, berg97, sana06, naze09}. 
This has been explained by the presence of shocks 
in the wind coming from instabilities due to its radiatively driven 
nature \citep{luwh80, lucy82, feld97a, feld97b}. 
For Wolf-Rayet stars, the situation is more complex: No detection of single WC star was reported in the X-ray domain \citep{oski03}; several single WN stars exhibit detectable X-ray emission \citep{skin10}, whereas other WN stars remain undetected in this band 
\citep[e.g.\ WR\,40,][]{goss05}.}. 
The variation of the former emission along the orbital cycle should 
provide information on the shape of the shock and its hydrodynamical nature. 
This variation is therefore an important source of knowledge of the relative strengths of the winds, hence, of the respective mass-loss rates.
The evolution with phase of the observed emission also varies because of the changing absorbing column along the line of sight which depends on the inclination of the system and the mass-loss rates. Therefore, the X-ray light curves of CWRs are of high diagnostic value for winds of massive stars.

In the above mentioned context, WR\,21a is a very interesting object. 
Known as an X-ray source since {\it{EINSTEIN}} observations, it was
the first Wolf-Rayet star discovered thanks to its X-ray emission 
\citep{mebe94, mere94}. Follow-up optical studies \citep{niem06,niem08,tram16}
showed that WR\,21a is actually a WN5+O3 binary system with a 
31.7\,d orbit.
The suggested mass for the primary WN star is $\sim$100\,M$_{\sun}$, 
making it one of the few examples of very high-mass stars. 
In order to deepen our knowledge of WR\,21a, we acquired four X-ray 
observations with the \xmm\ facility. Our aim was to obtain
an X-ray light curve for this supposed colliding-wind system as well as
X-ray spectra to interpret the behaviour of the collision zone 
along the orbital cycle. By adding archival data, we present 
a first interpretation of the X-ray light curve of this outstanding 
massive binary system. 
Section~2 contains the description of the observations and of 
the relevant reduction processes. 
Section~3 presents detailed information on the system, whilst Section~4 yields 
our analysis of the X-ray data. A discussion of the results is provided in 
Section~5 whilst we summarise and conclude in Section~6.

\section{Observations and data reduction} 
\subsection{Optical spectroscopy}
As a support to the \xmm\ observations, we acquired a few high resolution
spectra of WR\,21a with the FEROS spectrograph \citep{kauf99} linked 
to the ESO/MPG 2.2m telescope 
at the European Southern Observatory at La Silla (Chile).
Three spectra were secured in a run in 2006 
(ObsID = 076.D-0294, PI E.\ Gosset) 
and three in June 2013 (ObsID = 091.D-0622, PI E.\ Gosset). 
The latter spectra are of particular importance since they are 
contemporaneous with the \xmm\ pointings. 
A journal of the observations is provided in 
Table~\ref{optjournal}. 

The FEROS instrument provides 39 orders covering the entire 
optical wavelength domain,
going from 3800 to 9200\AA\ with a resolving power of 48000. 
The detector was a 2k $\times$ 4k EEV CCD with a pixel size of 15$\mu$m$\times$15$\mu$m. 
For the reduction process, we
used an improved version of the FEROS pipeline working under
the MIDAS environment \citep{sana06, mago12}.
The data normalisation was then performed by fitting polynomials 
of degree 4 - 5 to carefully chosen continuum windows. 
We mainly worked on the individual orders, but the regions around
the \ion{Si}{iv} $\lambda\lambda$4089, 4116,
\ion{He}{ii} $\lambda$4686, and H$\alpha$ emission lines were 
normalised on the merged spectrum as these lines appear at the 
limit between two orders.
Despite a one hour exposure time, the spectra have a S/N ratio 
of about 50-100 at the best, because of the 
faintness of the target. 

\begin{table}
\centering
\caption{Journal of the optical observations. Mid-exposure phases were calculated with the ephemeris of \citet{tram16}.}
\label{optjournal}
\begin{tabular}{lcccc}
\hline\hline
Date & ObsID & HJD (2400000+) & $\phi$\\
\hline
02-03-2006 & 076.D-0294 & 53796.553 & 0.543 \\
03-03-2006 & 076.D-0294 & 53797.579 & 0.575 \\
06-03-2006 & 076.D-0294 & 53800.706 & 0.674 \\
23-06-2013 & 091.D-0622 & 56467.501 & 0.853 \\
24-06-2013 & 091.D-0622 & 56468.497 & 0.885 \\
25-06-2013 & 091.D-0622 & 56469.506 & 0.917 \\
\hline
\end{tabular}
\end{table}
\subsection{X-ray observations}
\subsubsection{\xmm }
WR\,21a was observed four times with \xmm\ between mid-June 2013 
and July 2013 (orbits 2475, 2496, 2497 and 2497; see Table~\ref{journal}) 
in the framework of the programme 072419
(PI E.\ Gosset). The X-ray observations were made in the full-frame 
mode and the medium filter was used to reject optical/UV light. 
The data were reduced with SAS v13.5.0 using 
calibration files available in mid-October 2014 and following the recommendations of the \xmm\ team\footnote{SAS threads, see \\ http://xmm.esac.esa.int/sas/current/documentation/threads/ }. Data 
were filtered for keeping only best-quality data ({\sc{pattern}} of 0--12 
for MOS and 0--4 for pn). A background flare affecting the end of the last observation was also cut. 
A source detection was performed on each EPIC dataset using the task {\it edetect\_chain} on the 0.4--2.0 (soft), 2.0--10.0 (hard), and 0.4--10.0\,keV (total) energy bands and for a log-likelihood of 10. This task searches for sources using a sliding box and determines the final source parameters from point spread function (PSF) fitting; the final count rates correspond to equivalent on-axis, full PSF count rates (Table~\ref{journal}). 

We then extracted EPIC spectra of WR\,21a using the task {\it{especget}} in circular regions of 30\arcsec\ radius (to avoid nearby sources) centred 
on the best-fit positions found for each observation. For the background, 
a circular region of the same size was chosen in a region devoid of 
sources and as close as possible to the target; its relative position 
with respect to the target is the same for all observations. Dedicated ARF and RMF response matrices, which are used to calibrate the flux and energy axes, respectively, were also calculated by this task. EPIC spectra 
were grouped, with {\it{specgroup}}, to obtain an oversampling factor 
of five and to ensure that a minimum signal-to-noise ratio of 3 
(i.e.\ a minimum of 10 counts) was reached in each spectral bin 
of the background-corrected spectra. 

Light curves of WR\,21a were extracted, for time bins of 200\,s and 1\,ks, 
in the same regions as the spectra and in the same energy bands as 
the source detection. They were further processed by the task 
{\it epiclccorr}, which corrects for loss of photons due to 
vignetting, off-axis angle, or other problems such as bad pixels. 
In addition, to avoid very large errors and bad estimates of the 
count rates, we discarded bins displaying effective exposure time 
lower than 50\% of the time bin length. Our previous experience 
with \xmm\ has shown us that including such bins degrades the results. 
As the background is much fainter than the source, in fact too faint to 
provide a meaningful analysis, three sets of light curves were produced 
and analysed individually: the raw source+background light curves, 
the background-corrected light curves of the source and the light curves 
of the sole background region. The results found for the raw and background-corrected light curves of the source are indistinguishable. 

\subsubsection{{\it{Swift}}}
WR\,21a was observed 198 times by {\it{Swift}} in October-November 2013, in 
June and October-November 2014 as well as in January 2015 (see Table~\ref{journal}). These data were retrieved 
from the HEASARC archive centre.

XRT data were processed locally using the XRT pipeline of HEASOFT v6.16 
with calibrations available in mid-October 2014. Corrected count rates 
in the same energy bands as \xmm\ were obtained for each observation 
from the UK on-line tool\footnote{http://www.swift.ac.uk/user\_objects/} 
(Table~\ref{journal}), which also provided the best-fit position 
for the full dataset (10$^\mathrm{h}$ 25$^\mathrm{m}$  56\fs 48, 
\mbox{--57\degr 48\arcmin 43\farcs 5}, similar to Simbad's value). 
This position was used to extract the 
source spectra within Xselect in a circular region of 47\arcsec\ 
radius (as recommended by the {\it{Swift}} team). 
They were binned using {\it grppha} in a similar manner as the \xmm\ spectra. Following the recommendations 
of the {\it{Swift}} team, a background region as large as possible was chosen, 
i.e.\ an annulus of outer radius 130\arcsec. 
The most recent RMF matrix from the calibration database was used 
whilst specific ARF response matrices were 
calculated for each dataset using {\it xrtmkarf}, and considering the 
associated exposure map. In about half (112 out of the 198) of the 
exposures, WR\,21a displays few raw counts, which renders 
spectral fitting unreliable. Thus we only present the count rates 
of these exposures. 

\begin{figure}
\includegraphics[width=8.5cm]{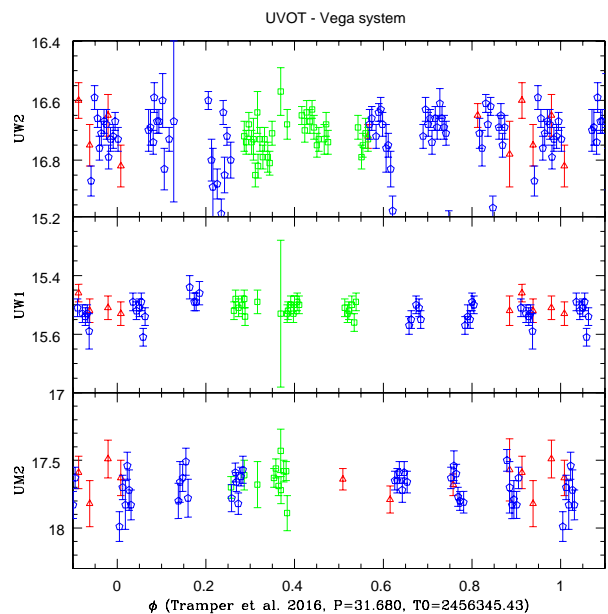}
\caption{UVOT magnitudes of WR\,21a as a function of phase. 
Data from 2013 are shown with empty red triangles, data from 2014 
with empty green squares, and data from 2015 with empty blue pentagons.}
\label{uvot}
\end{figure}

For UVOT data, we defined a source region centred on the same 
coordinates\footnote{There is however a coordinate shift for eight 
UVOT datasets but the relative positions of source and background were preserved in those cases.} but with 5\arcsec\ radius, as recommended by the 
{\it{Swift}} team. 
Because of the straylight UV emission from a nearby Be star, HD\,90578, 
a background region as close to WR\,21a as possible was chosen to obtain 
a representative background. It also avoids other nearby, faint UV sources; 
this background region is centred on 10$^\mathrm{h}$ 25$^\mathrm{m}$ 58\fs 372,
--57\degr\ 48\arcmin\ 32\farcs 94 and has a 10\farcs 75
radius; it was used for all observations except 00032960033 
(where spikes from the Be star contaminate this region, forcing us to shift its centre to 10$^\mathrm{h}$ 25$^\mathrm{m}$ 54\fs 846, 
--57\degr\ 49\arcmin\ 00\farcs 49). Vega magnitudes were then derived 
via the task {\it{uvotsource}}. They are shown in Fig.~\ref{uvot}; 
no significant variation is detected within the limits of the noise. 
The absence of strong variations and of eclipses, in particular, are confirmed in the optical range (Gosset \& Manfroid, in prep.).

\subsubsection{\it{Chandra}}

The {\it{Chandra}} X-ray facility observed serendipitously WR\,21a with ACIS-I in April 2008. The system appears far off-axis, near a CCD edge. Consequently, the PSF is heavily distorted and the counts are spread over a large area, which avoids the pile-up of the source. The spectrum of WR\,21a was extracted in a circle of radius 24\farcs 3 centred on its Simbad coordinates, that of the background in a nearby region of the same size and devoid of sources. Dedicated ARF and RMF response matrices were calculated via 
{\it{specextract}} (CIAO v4.7 and CALDB v4.6.9). 
The X-ray spectrum was binned in a similar manner to the \xmm\ spectra. Count rates in the same energy bands as for \xmm\ and {\it{Swift}} data were derived from the ungrouped spectrum using Xspec (see Table~\ref{journal}) - they are thus not equivalent on-axis values.

\section{The WR\,21a system}

The star now known as WR\,21a was first detected as an H$\alpha$-emitter, appearing in the \citet{thee66} catalogue under the name THA\,35-II-042 and in the early-type emission-line star catalogue of \citet{wack70} as Wack\,2134. 
The star is situated to the east of
Westerlund\,2 \citep[see e.g.\ figure~3 of][]{roma11} but its relation to that
cluster remains unknown. In fact, the exact distance to the star is currently unknown, and even a precise $V$ magnitude is lacking for WR\,21a.
\citet{wack70} provides a value $V$~=~12.8 whereas the more recent
UCAC4 gives $V$~=~12.67 \citep{zach13}. 
In our spectra,
we note that the interstellar \ion{Na}{i} D lines have components
with velocities from --12 km\,s$^{-1}$ to +10 km\,s$^{-1}$. Adopting the
Galactic rotation law of \citet{fich89}, we then deduce a kinematical
distance of about 5--5.4\,kpc which is in between the small ($\sim$2.6\,kpc, \citealt{asce07,acke11}) and large ($\sim$8\,kpc, \citealt{rauw11}) distance values of the Westerlund 2 cluster, but similar to the middle determination of \citet{fuku09} and \citet{varg13}. The interstellar \ion{K}{i} $\lambda$7699 and
CH $\lambda$4300 lines exhibit the local absorption component up to +10 km\,s$^{-1}$ on the red side but display nothing on the blue side.

\begin{figure}
\includegraphics[width=8.5cm]{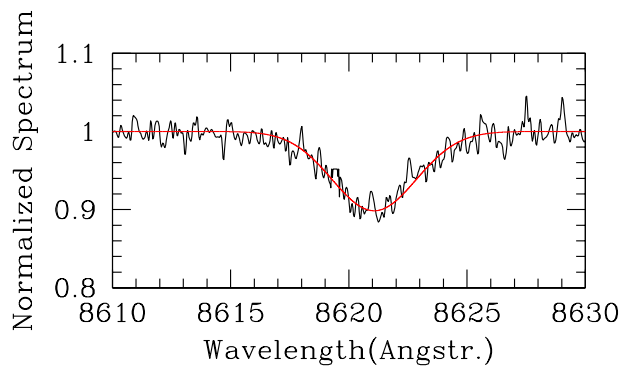}
\caption{Mean FEROS spectrum of WR\,21a around the DIB at 8621\AA. The
red curve gives the fit by a Gaussian function of dispersion
$\sigma$~=~1.78 \AA . }
\label{figdib}
\end{figure}

Figure~\ref{figdib} shows the spectrum of WR\,21a
in the region of the diffuse interstellar band (DIB) at 8621\AA. We measured an equivalent width
of 0.42\AA\ for this feature. Following the calibration of
\citet{wasa07}, this corresponds to an excess $E(B-V)$=1.8 in good agreement 
with the conclusion of $E(B-V)$=1.48--1.9 from \citet{cara89}. Another estimate can be calculated from the
$JHK$ magnitudes of the system from 2MASS data. Correcting these
magnitudes to the standard system, we derived
$(J-H)$\,=\,0.664 and $(H-K)$\,=\,0.343. Dereddening the $(J-H)$ colour
to the typical colour of massive O-stars, we deduced an equivalent excess 
$E(B-V)$\,=\,1.6 to 2.0. The same exercise was made considering 
both the $(J-H)$ and $(H-K)$
intrinsic colours of a WN5h star as calibrated by
\citet{rocr15}, which yields  
$E(B-V)$\,=\,1.7 to 1.9, further supporting the above-mentioned 
value of 1.8 for this excess. We thus adopt this value
for the present work. Using the calibration of 
\citet{bohl78} and \citet{gud12}, this colour excess corresponds to an equivalent hydrogen column density
of 10$^{22}$ cm$^{-2}$. 

\subsection{WR\,21a as a high-energy source}
WR\,21a was tentatively identified as the optical counterpart
of the X-ray source 1E\,1024.0--5732 from the {\it{EINSTEIN Galactic Plane Survey}} \citep{hegr84}. It was then suggested that this X-ray source was not coronal in nature, as in sun-like stars. In addition, this X-ray source was amongst a small group 
of objects present in the error box of the $\gamma$-ray source 2CG\,284--00 \citep{cara83, gold87}.
A few years later, \citet{cara89} confirmed the association 
of the X-ray source with the early-type star but also reported 
the detection of a 60\,ms pulsation in the X-ray emission
that they considered reminiscent of a pulsar. 
They thus concluded that it was a O+neutron star binary. 
However, \citet{diet90} discarded the presence of any pulsation in the visible
domain and \citet{beme94} were also unable to find support for the 60\,ms pulsations in {\it{ROSAT}} observations. In parallel, observations in the visible domain instead suggested that the optical spectrum was of the 
type WN6 (or WN5) with a possible companion.
This made Wack\,2134 the first Wolf-Rayet star discovered thanks to 
its X-ray emission \citep{mebe94, mere94, mere95}.  

Using the {\it{Rossi X-ray Timing Explorer}}, \citet{reig99} analysed the
3--15\,keV spectrum of 1E\,1024.0--5732. 
He concluded that it was a rather soft 
source (hence not an accretor), clearly favouring a colliding wind 
scenario rather than the HMXB scenario. He further refuted the presence 
of any rapid fluctuations but mentioned that the source is probably variable 
on timescales of years. WR\,21a has also been observed with the 
{\it{ASCA Gas Imaging Spectrometer}} and appears as the faint and low-variability source AX\,J1025.9--5749 in the catalogue of
\citet{robe01}. In addition, WR\,21a is, in the radio domain, 
a possible non-thermal emitter \citep{bena05, dera13},
which is another possible characteristic of colliding winds.

\begin{figure}
\includegraphics[width=8.5cm]{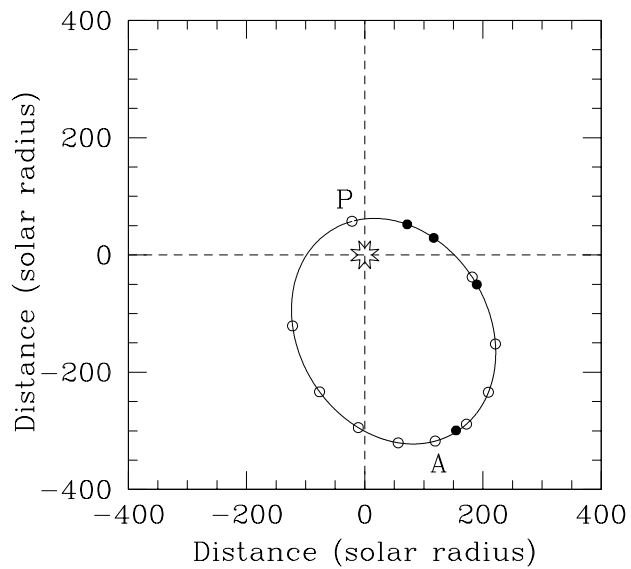}
\caption{Relative orbit of the O-star in WR\,21a around the Wolf-Rayet 
primary star. The orbit is projected on the plane defined by the line of sight (vertical dashed line) and the line of nodes (horizontal dashed line). 
This was calculated according to the orbital 
solution of \citet{tram16} with an assumed inclination of 58\fdg 8. 
Both axes are in units of the solar radius; the Earth is located to the 
bottom at $y = -\infty$ on the ordinate axis. The orbital motion is counterclockwise. Open circles indicate the position of the O-star between phases 0 and 1 by step of 0.1. Periastron (P) occurs at phase $\phi$~=~0.0 and apastron (A) at $\phi$~=~0.5. The filled circles indicate
the binary configuration at each of the \xmm\ observations. 
The passages through
the lines of nodes (horizontal dashed line) occur at $\phi$~=~0.0335 and $\phi$~=~0.928, whereas the conjunctions (intersections between the orbit and the vertical dashed line) occur at $\phi$~=~0.316 and $\phi$~=~0.9935.
}
\label{orbproj}
\end{figure}

\subsection{WR\,21a as a binary system}

Despite the suspicion of binarity from the high-energy studies, the characteristics of WR\,21a remained poorly known until recently. The first detailed analyses by \citet{niem06} and \citet{niem08}
definitively proved that WR\,21a is actually a binary system with a period of 31.673\,d and a large eccentricity.
The absence of \ion{He}{i} lines for the O companion indicated an early 
O3--O4 star, whilst the spectral type of the WR was estimated to WN6. Quite recently, the system has been densely monitored with 
X-Shooter at ESO's VLT by \citet{tram16}. They broadly confirmed the previous results and improved the orbital solution. We adopt this latter
solution (Table~\ref{orb} and Fig.~\ref{orbproj}). 
The spectral types were revised to O3/WN5h+O3Vz((f$^*$)).

\begin{table}
\centering
\caption{Physical parameters of WR\,21a. The first part of the Table 
yields the orbital solution from \citet{tram16} and the semimajor axis $a$ derived from it. The second part of the Table gives the adopted 
stellar parameters (see Sect.~3.2 for details).}
\label{orb}
\begin{tabular}{lcc}
\hline\hline
\multicolumn{1}{l}{Orbital Parameters (Units)} & 
\multicolumn{2}{c}{Value}\\
& Primary & Secondary \\
\hline
$P$ (d) & \multicolumn{2}{c}{31.680 $\pm$ 0.013} \\
$e$ & \multicolumn{2}{c}{0.694 $\pm$ 0.005} \\
$q$ (Prim./Sec.) & \multicolumn{2}{c}{1.782 $\pm$ 0.030} \\
$T_0$ (HJD-2450000) & \multicolumn{2}{c}{6345.43 $\pm$ 0.32} \\
$\omega$ (\degr) & \multicolumn{2}{c}{287.8 $\pm$ 1.2} \\
$K$ (km s$^{-1}$) & 157.0 $\pm$ 2.3 & 279.8 $\pm$ 6.2 \\
$\gamma$ (km s$^{-1}$) & --32.8 $\pm$ 1.7 & 32.8 $\pm$ 2.9 \\
$a \sin(i)$ (R$_{\sun}$) & 70.8 $\pm$ 2.9 & 126.1 $\pm$ 1.1 \\
$M \sin^3(i)$ (M$_{\sun}$) & 65.3 $\pm$ 5.6 & 36.6 $\pm$ 1.9 \\
\hline
\multicolumn{1}{l}{Stellar Parameters (Units)} & 
\multicolumn{2}{c}{Value}\\
& Primary & Secondary \\
\hline
Spectral types & WN5h & O3V  \\
$\log$ $\dot{M}$ (M$_{\sun}$\,yr$^{-1}$) & --4.5 & --5.64 \\
$v_{\infty}$ (km s$^{-1}$) & 2000 & 3800 \\
$R$ (R$_{\sun}$) & 12.0 & 13.84 \\
\hline
\end{tabular}
\end{table}
The orbital solution indicates rather large minimum masses 
($M \sin^3 \, i$ of 65.3 M$_{\sun}$ for the WN star and
36.6 M$_{\sun}$ for the O companion), putting WR\,21a in the list of very massive systems with a WNLh primary
\citep[like WR\,22, WR\,25, WR\,20a, and WR\,29; see respectively][]{rauw96, gago06, rauw05, gafe09}. 
The inclination is basically unknown, but
if a mass of 58 M$_{\sun}$, typical of O3V stars 
\citep{mart05} is attributed to the O-type companion, an inclination $i$\,=\,58\fdg 8 is deduced, leading to a high mass of 104 M$_{\sun}$ for the WN5h star \citep{tram16}. WR\,21a would thus be one of the rare examples of the most massive stars ($M\gtrsim$ 100\,M$_{\sun}$), underlining its interest. 

\begin{figure}
\centering
\includegraphics[width=9cm]{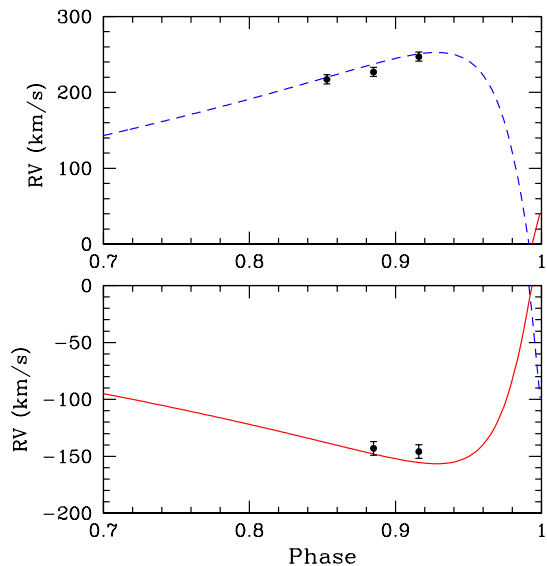}
\caption{The radial velocity curves for both components in WR\,21a from the orbital solution of \citet{tram16}. The dashed blue and continuous red lines provide the curves for the secondary and primary, respectively. 
The dots represent the RVs measured on the 2013 FEROS spectra acquired contemporaneously 
with the \xmm\ data. The agreement with the pre-established orbital
solution is good. The error-bars represent 1-$\sigma$
standard deviation.}
\label{figrvab}
\end{figure}

The ephemeris derived by \citet{tram16} were used for deriving the phases of the X-ray observations (see Table~\ref{journal}). They are precise: The error on the reference time of periastron amounts to 0.32\,d, inducing a possible error on the phase of 0.01;
the error on the period (0.013\,d) turns out, over five cycles, to an error on the phase of 0.002 only. Nevertheless, to be sure that the phases derived for our \xmm\ data are correct, we measured the radial velocities of the 
\ion{He}{ii} $\lambda$5412 line of both objects on the contemporaneous optical spectra. Comparing them with the predictions from the orbital solution of \citet[see also Fig.~\ref{figrvab}]{tram16}, we found that the error on the phases is not larger than 0.01. Uncertainties should be slightly larger for {\it{Swift}} observations taken in 2014 and 2015 as well as for the 
{\it{Chandra}} 2008 spectrum, as the reference time of periastron in \citet{tram16} is in 2013 and there are about 11-12 cycles per year, but errors on the phases should nevertheless remain below 0.05 for {\it{Swift}} and 0.12 for {\it{Chandra}}.
This is confirmed by the near-perfect reproduction of the X-ray light curve with time (see next section).

\begin{figure*}
\includegraphics[width=8.5cm, bb=40 320 590 720,clip]{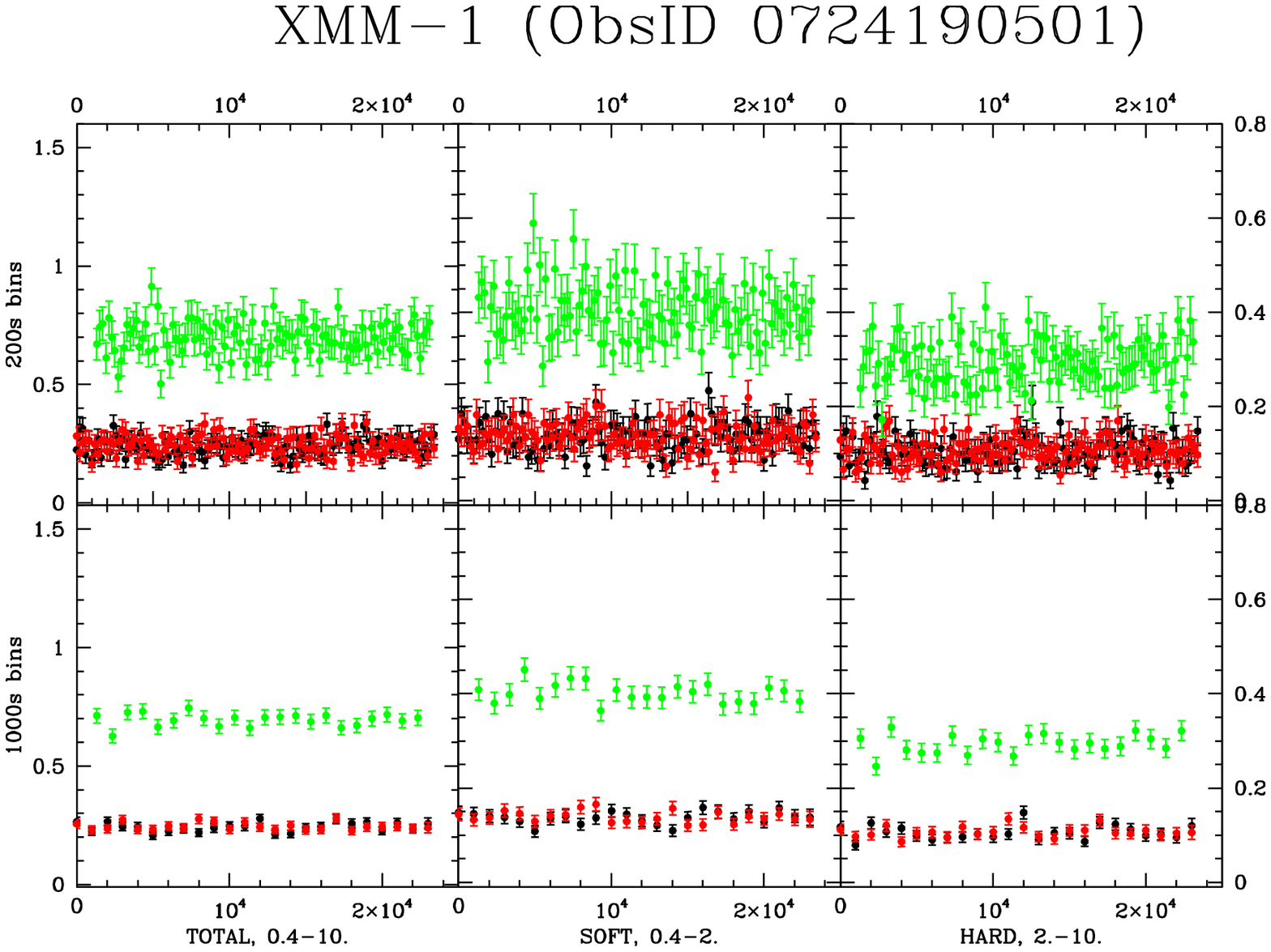}
\includegraphics[width=8.5cm, bb=40 320 590 720,clip]{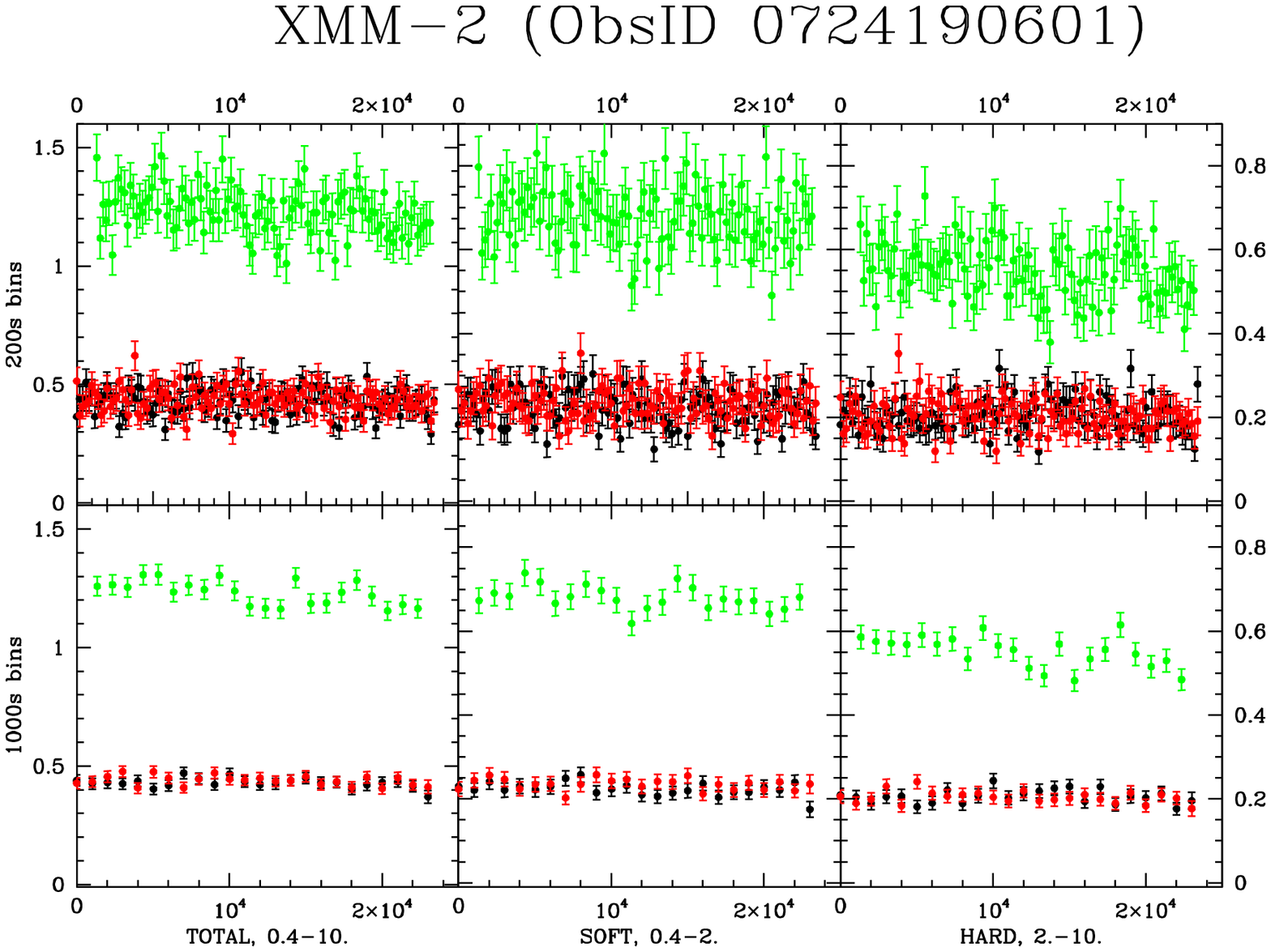}
\includegraphics[width=8.5cm, bb=40 320 590 720,clip]{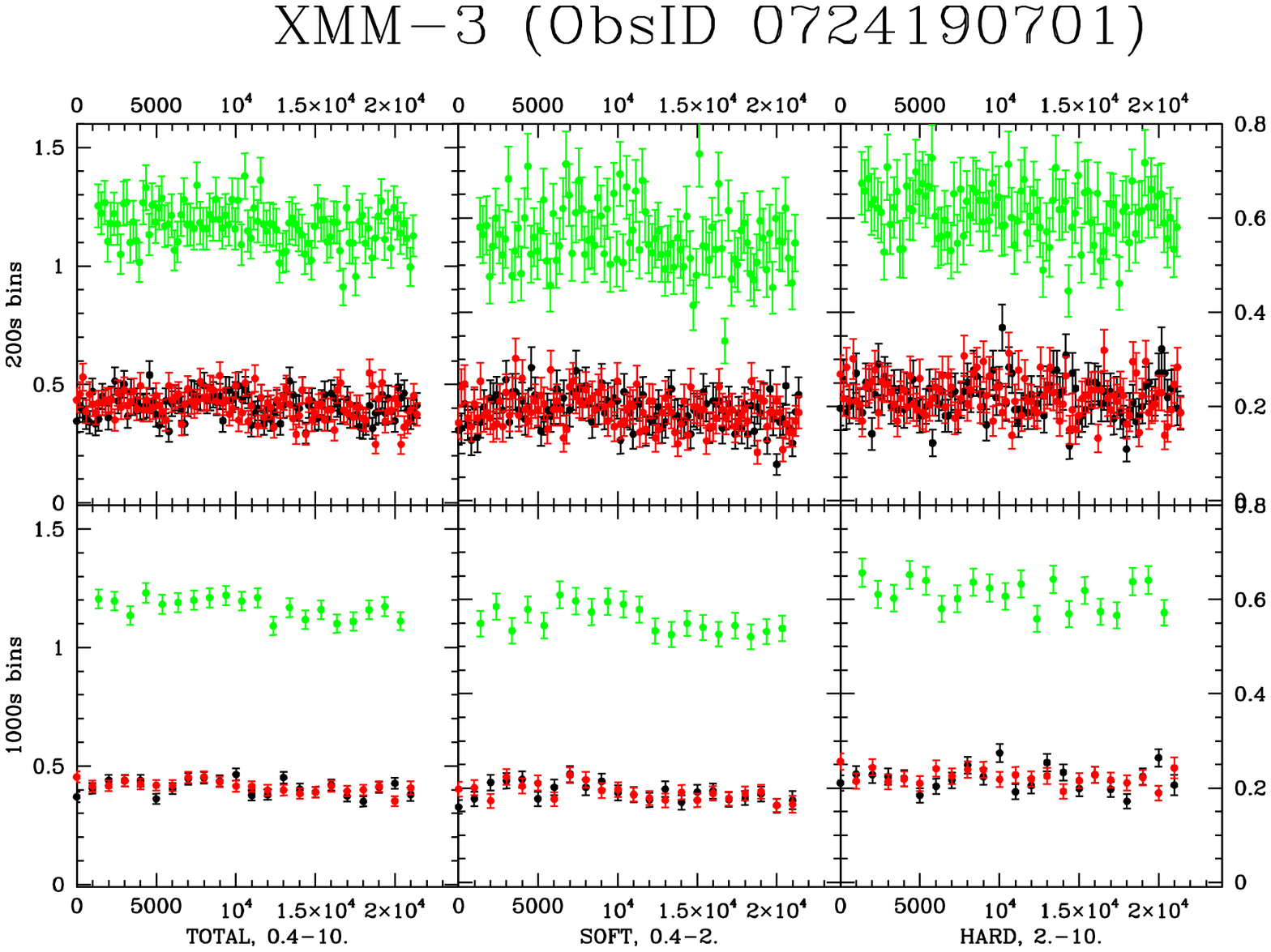}
\hfill
\includegraphics[width=8.5cm, bb=40 320 590 720,clip]{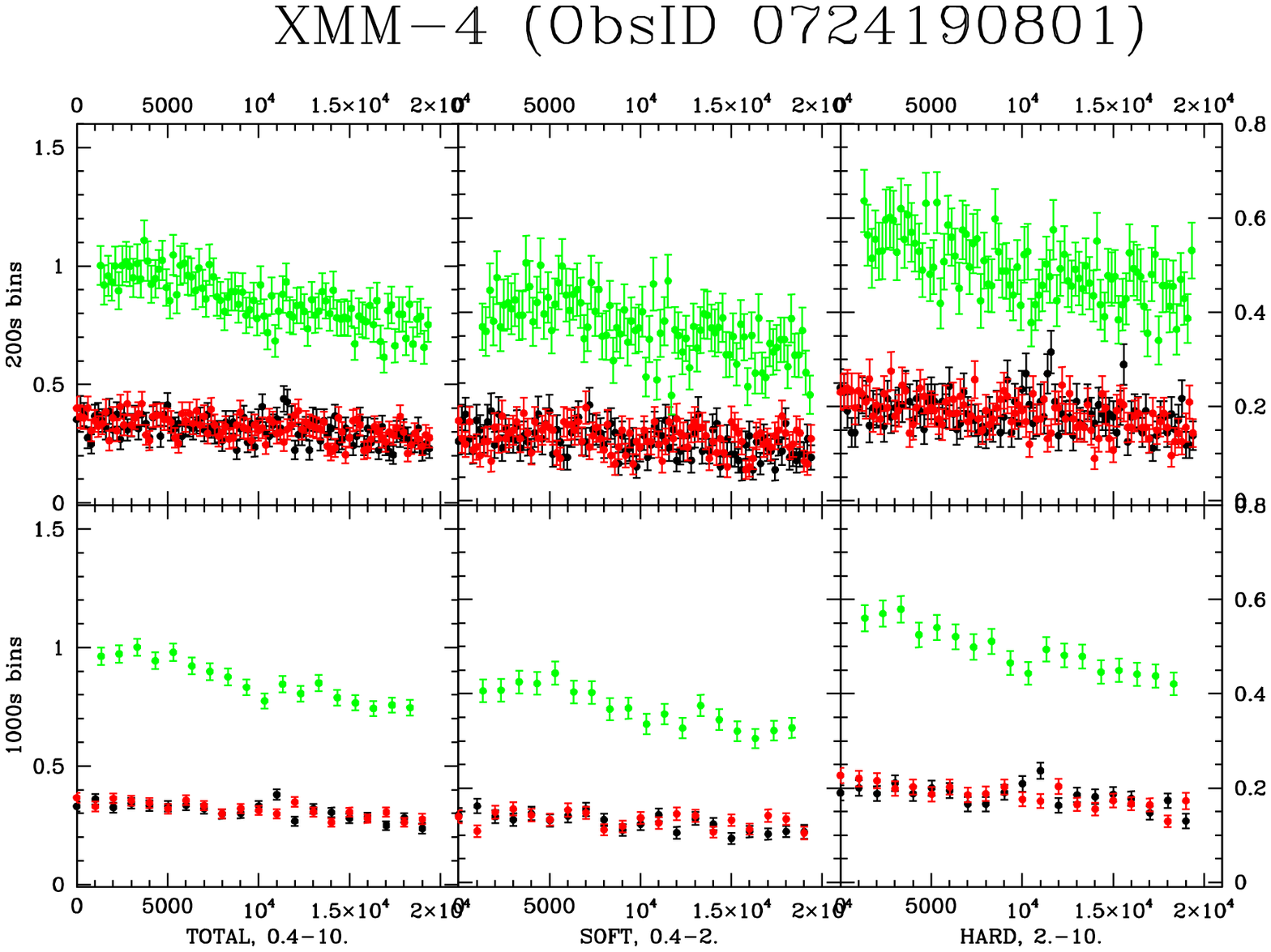}
\caption{Light curves of the four \xmm\ observations (pn in green, MOS1 in black, MOS2 in red). For each exposure, the top (resp. bottom) panels show the light curves with 200\,s (resp.\ 1000\,s), whilst the left, central, and right panels display the total, soft, and hard light curves, respectively. The ordinates
are in count/s: on the left side for the total light curves, on the right
side for the soft and hard light curves.}
\label{lcxmm}
\end{figure*}

Despite these previous studies, the stellar and wind parameters are still largely unknown. This is notably owing to the lack of a detailed study 
of the WN spectrum with stellar atmosphere codes
such as e.g. CMFGEN \citep[Co-Moving Frame GENeral, ][]{hil98}. 
The second part of Table~\ref{orb} thus yields parameters, 
such as star radii and mass-loss rates,
which are inspired from several studies of similar objects.
Considering the primary star, we used the analysis of equivalent
WN5-6(h) stars by \citet{crow95} and \citet{hama06} as well as the
parameters of the WN7ha star WR\,22 \citep{goss09, graf08} and of the
WN6ha star WR\,20a \citep{rauw05}. Parameters for the O companion come
from the calibration of \citet{mart05} and of
\citet{muij12}.
Whilst they should certainly not be taken at face value, these parameters can be considered as representative, helping to get a first idea of the nature of the wind-wind collision; these parameters suggest that the wind momentum of the WR star is overwhelming the one of the O-star. 
Under the hypothesis of instantaneously accelerated winds, we expect the apex of the collision to be on the binary axis at 73\% of the system separation
from the WR, i.e.\ close to the O-star. 
If we further consider some radiative braking acting on the WR wind, the apex moves away from the O-star.
Using the formalism of \citet[see their equation 4]{gayl97}, the apex 
might shift to 58\% of the separation if we adopt a 
maximum value of about two for the reflection factor $S$. 
If instead we consider a radiatively accelerated wind obeying a classical $\beta$-velocity law (e.g.\, $\beta$=0.8 for both stars), we still find a value of 73\% (neglecting the braking) for the apex position during the major part of the orbit
(roughly from phase 0.2 to 0.8). Around periastron, it appears that
the O-star wind could have difficulty in supporting the WR wind 
with, hence, a possible crash on the surface of the O-star. 
However, this possibility is attenuated if braking is considered.
In particular, the maximum value of $S$ corresponds to
an ability for the O-star to fully sustain the WN wind, restoring
the position of the apex at 57\%.

The nature of the collision can be derived from the value of the 
index $\chi$, which represents the ratio of the 
typical cooling time over the escaping time \citep{stev92}. 
For solar abundances, we have $\chi=\frac{t_{\mathrm{cool}}}{t_{\mathrm{esc}}}=\frac{v^4_{1000}\times d_{12}}{\dot M_{-7}}$, where $v_{1000}$ is the pre-shock velocity in units of 1000\,km\,s$^{-1}$, $d_{12}$ is the star-apex separation in units of $10^7$\,km, and $\dot M_{-7}$ is the mass-loss rate in units of $10^{-7}$\,M$_{\odot}$\,yr$^{-1}$. For our adopted values, it is in the range 0.3--40 for the O-star (depending on the actual location of the apex, on the considered phase, and on the actual 
inclination of the system) but it is always less than 0.35 for the WR (taking into account the WN abundances).
The post-shock O wind thus is in an intermediate state between radiative and adiabatic behaviours, but on the adiabatic side, whereas
the shocked WR wind is at best in an intermediate state, 
but on the rapidly cooling side.
Therefore, one could expect a weakly varying hard component and 
instabilities a little more developed than in the case of
WR\,22 \citep{park11}. We are now going to check these ideas by analysing the results of the \xmm\ observations.
\section{The X-ray emission of WR\,21a and its CWR}

\begin{table*}
\centering
\caption{Significance levels (in \%) associated with $\chi^2$ tests for constancy of the background-corrected \xmm\ light curves with 1\,ks bins. A value lower than 1\% is considered as a detection of significant variability. S, H, and T refer to the 0.4--2.0\,keV, 2.0--10.0\,keV and 0.4--10.0\,keV energy bands, respectively.}
\label{chi2}
\begin{tabular}{l|ccc|ccc|ccc|ccc}
\hline\hline
Inst. & \multicolumn{3}{c|}{XMM-1} & \multicolumn{3}{c|}{XMM-2} & \multicolumn{3}{c|}{XMM-3}  & \multicolumn{3}{c}{XMM-4}\\
& T & S & H & T & S & H & T & S & H & T & S & H\\
\hline
pn   & 69 & 71 & 34 & 3 & 62 & $<1$ & 23 & 42 & 21 & $<1$ & $<1$ & $<1$ \\ 
MOS1 & 14 & 39 & 3 & 89 & 64 & 41 & $<1$ & 10 & $<1$ & $<1$ & $<1$ & $<1$ \\
MOS2 & 77 & 71 & 56 & 70 & 86 & 73 & 27 & 19 & 54 & $<1$ & 4 & $<1$ \\
\hline
\end{tabular}
\end{table*}

\begin{figure*}
\includegraphics[width=8.5cm]{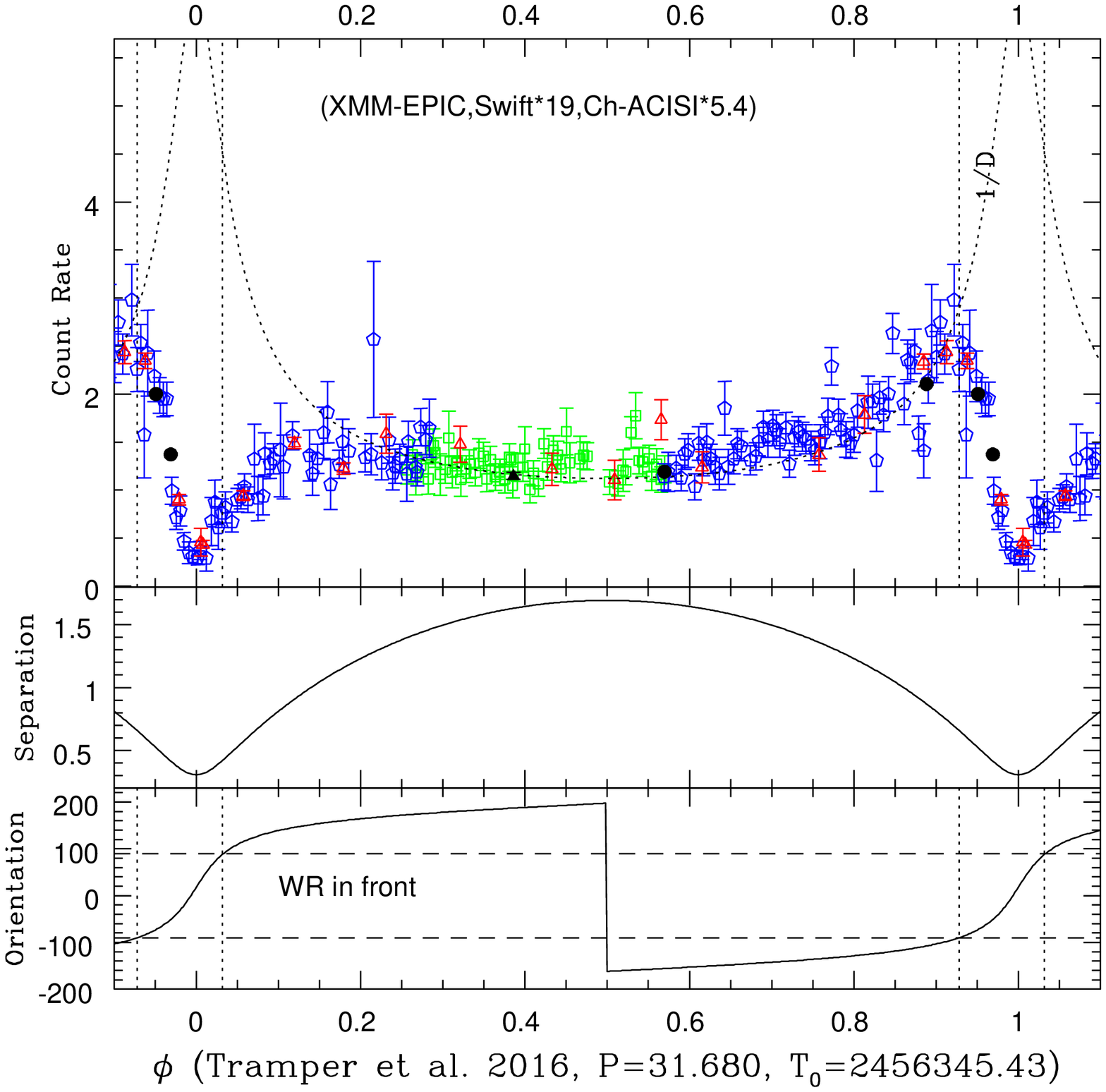}
\hfill
\includegraphics[width=8.5cm]{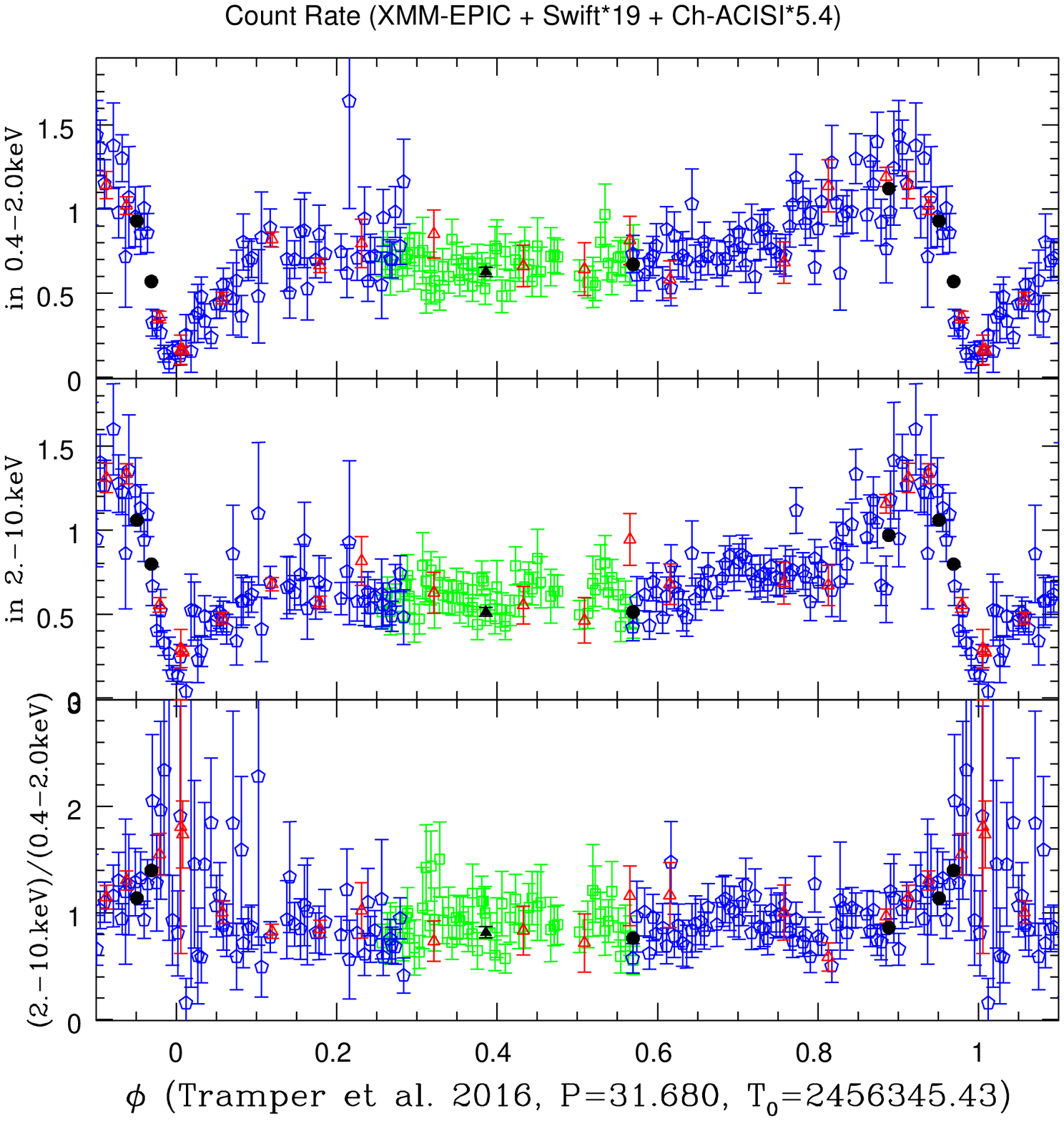}
\caption{Variation of the count rates and hardness ratio with orbital 
phase, for \xmm\ (in black dots), {\it{Chandra}} (filled black triangle), 
and {\it{Swift}} data (empty red triangles for 2013, empty green squares 
for 2014, and empty blue pentagons for 2015). For all energy bands, {\it{Chandra}} and {\it{Swift}} count rates and their errors were 
multiplied by 5.4 and 19, respectively, to clarify the trends, and no further treatment was made for adjustment. To help compare with physical parameters, the bottom left panels provide the orbital separation (in units of the semimajor axis $a$) as well as a position angle defined as zero when the WR star is in front and 180$\degr$ when the O-star is in front. The vertical dotted lines in the left panels thus correspond to the quadratures and the other dotted line to an arbitrary $1/D$ variation (not fitted to the data); periastron occurs at $\phi=0.0$.}
\label{lcall}
\end{figure*}

\subsection{The light curves}
The good sensitivity of \xmm\ allows us to search for variations within the exposures (Fig.~\ref{lcxmm}). We note that differences are expected between instruments as pn and MOS do not have the same spectral sensitivity and do not receive the same amount of photons; about half of the flux in front of MOS telescopes is redirected into the reflexion grating spectrometer (RGS) leading to larger noise in MOS data. Besides, noise makes light curves recorded even by twin-like instruments that are not exactly identical (for more discussion and examples, see \citealt{naze13}). All these problems can however be overcome by cautious statistical testing. As for $\zeta$\,Pup \citep{naze13}, the same set of tests was applied to all cases. We first performed a $\chi^2$ test for three 
different null hypotheses (constancy, see e.g. Table~\ref{chi2}; linear variation; quadratic 
variation), and further compared the improvement of the $\chi^2$ when 
increasing the number of parameters in the model (e.g. linear trend vs constancy) by means of Snedecor F tests 
\citep[nested models, see Sect.~12.2.5 in][]{lind76}. Adopting a significance level of 1\%, we found that WR\,21a has not significantly varied during the first \xmm\ observation, but that the hard pn light curve with 1\,ks time bins in the second \xmm\ observation is found to be significantly variable and significantly better fitted by a linear relation. In the third \xmm\ observation, the pn and MOS2 light curves in the total energy band are significantly better fitted by linear or quadratic relations than by a constant whilst the MOS1 light curve appears significantly variable. Finally, in the last observation, WR\,21a always appears significantly variable and better fitted by linear or quadratic relations. Indeed, a decrease of the count rate is obvious in the last two \xmm\ observations (Fig.~\ref{lcxmm}), although it is shallower for the former one of these last two.

\begin{figure*}
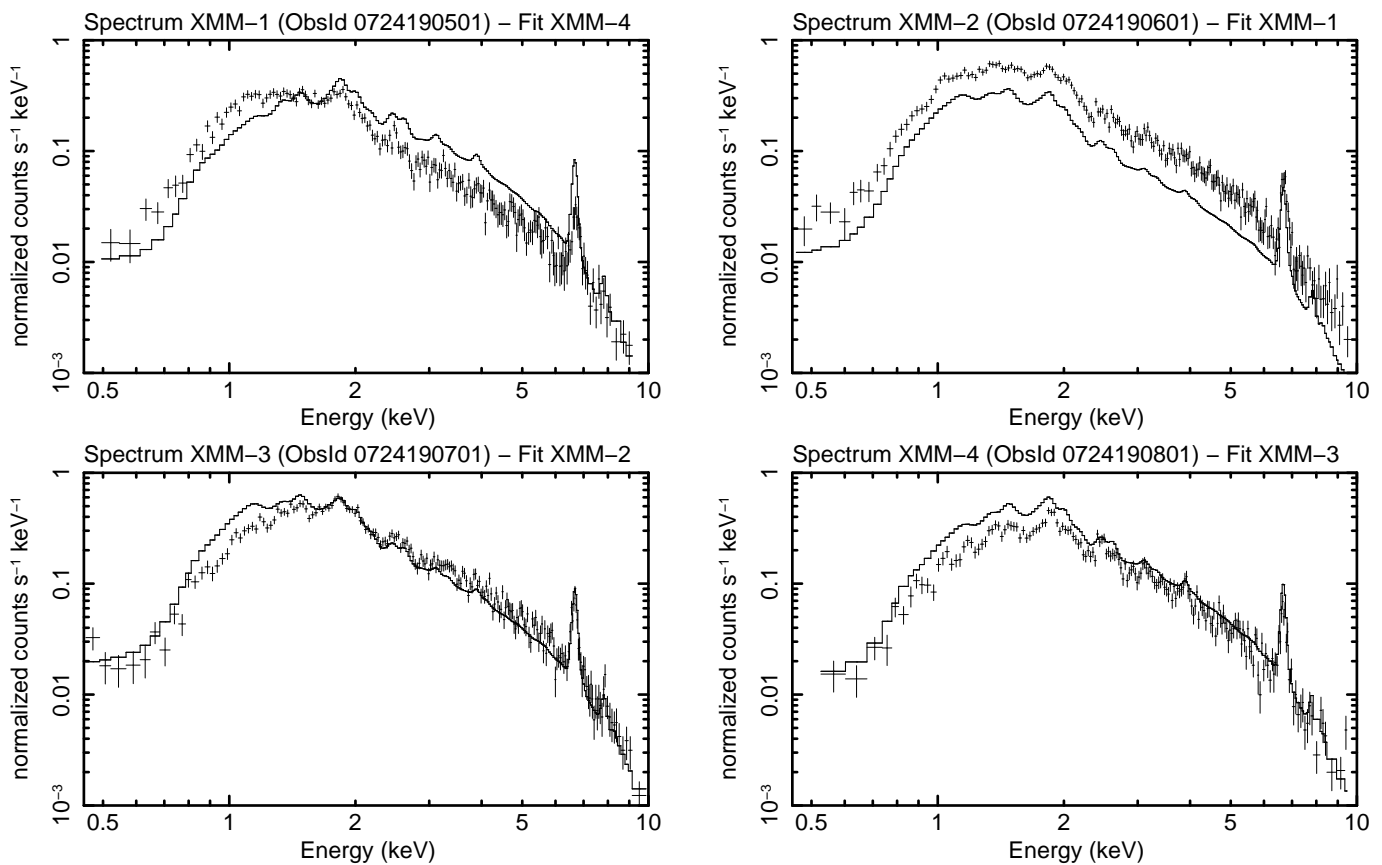

\includegraphics[width=5.7cm,angle=270,bb=280 40 570 510,clip]{fig2475new.ps}
\includegraphics[width=5.7cm,angle=270,bb=280 40 570 510,clip]{fig2496new.ps}
\includegraphics[width=5.7cm,angle=270,bb=280 40 570 510,clip]{fig2497anew.ps}
\includegraphics[width=5.7cm,angle=270,bb=280 40 570 510,clip]{fig2497bnew.ps}
\caption{Variation of the shape of the XMM-pn spectra with time, each panel
showing an individual spectrum. 
In each panel, we also superimposed the best-fit model 
(absorbed 2T model with individual absorption columns)
of the preceding (in phase) pn spectrum (see Sect.~4.2).}
\label{spec}
\end{figure*}

To put these results into context, we also looked at the global light curves, i.e. light curves combining \xmm, {\it{Chandra}}, and {\it{Swift}} data, with one point per exposure (Fig.~\ref{lcall})\footnote{In this figure, {\it{Chandra}} and {\it{Swift}} count rates are multiplied by 5.4 and 19, respectively, whatever the energy band. This empirical correction does not rely on any theoretical assumption. It is only used to show in a simple way that all datasets agree throughout the whole orbit. However, the values of the applied factors agree well with what can be derived from simulations in Xspec (using {\it fakeit} and models of Table~\ref{spectralfits}) and in PiMMS (for simpler emission models).}. Variations recorded by these observatories agree well, even if they are several cycles apart. Moreover, it is obvious that the last two \xmm\ observations were taken when WR\,21a becomes fainter, explaining the above results. In fact, the brightness of WR\,21a does not change much from $\phi=0.2$ to 0.7, but important variations are seen between $\phi=0.8$ and 1.1, i.e. around periastron passage. First, the count rate increases, by about 80-100\% with a maximum before conjunction and periastron (at $\phi\sim 0.9$). Next, a sharp drop in the soft band is detected 
with a minimum reached slightly before $\phi\sim 0.0$, most probably at conjunction (which occurs at $\phi\sim\,0.9935$). 
Despite the fact that the hard band must be much less sensitive to absorption effects, a drop is also observed in this band. Moreover, the amplitude of the drop does not strongly depend on energy. The count rates at minimum are about seven times smaller than these at $\phi=0.2-0.7$. In parallel, the hardness ratio increases from $\phi=0.95$ up to $\phi\sim0.0$ but comes back to its original value after that. This suggests that the minimum
occurs slightly later in the hard band than in the soft band.
It may be noted that the behaviour of the light curve appears more complicated than a simple eclipse by the WR near $\phi\sim0.9935$. Also, there is clearly no eclipse when the O-star is in front (around $\phi=0.316$), as would be expected for large inclination systems (e.g. V444 Cyg, \citealt{loma15}).

Finally, the increase in the count rate before periastron appears to follow a $1/D$ trend (see dotted line in Fig.~\ref{lcall}), typical of adiabatic systems (where $\chi>>1$). In fact, this trend can actually be detected from $\phi=0.2$ to $\phi=0.9$. This point is discussed further below.

\subsection{The spectra}
\begin{table*}
\centering
\caption{Results of the spectral fits of \xmm\ observations above 3.0\,keV.}
\label{egspectralfits}
\begin{tabular}{lccccccc}
\hline\hline
\multicolumn{6}{l}{Model $powerlaw+gaussian$}\\
ID & $\phi$ & $\Gamma$ & $norm$ at 1\,keV & Position & Width ($\sigma$) & Line Flux & $\chi^2$ (dof) \\ 
&&& (10$^{-3}$\,ph.\,keV$^{-1}$\,cm$^{-2}$\,s$^{-1}$) & (keV) & ($10^{-2}$\,keV) & ($10^{-6}$\,ph.\,cm$^{-2}$\,s$^{-1}$) \\
\hline
1& 0.57  &2.71$\pm$0.07 &2.12$\pm$0.21 &6.717$\pm$0.015 &7.67$\pm$2.31 &9.04$\pm$1.03 &0.85 (234)\\
2& 0.89  &2.77$\pm$0.04 &4.51$\pm$0.32 &6.676$\pm$0.007 &0.19$\pm$0.96 &15.5$\pm$1.12 &1.00 (293)\\
3& 0.95  &2.85$\pm$0.04 &5.67$\pm$0.37 &6.671$\pm$0.010 &3.83$\pm$2.95 &18.4$\pm$1.80 &1.13 (291)\\
4& 0.97  &2.91$\pm$0.05 &5.03$\pm$0.38 &6.670$\pm$0.009 &5.67$\pm$1.53 &18.0$\pm$1.40 &1.25 (261)\\
\hline
\multicolumn{5}{l}{Model $apec$}\\
ID  & $\phi$ & $kT$ & $norm$ & $\chi^2$ (dof) \\
& & (keV) & ($10^{-3}$\,cm$^{-5}$) & \\
\hline
1&0.57  &3.19$\pm$0.13 &2.68$\pm$0.11 &1.08 (237)\\
2&0.89  &2.99$\pm$0.08 &5.59$\pm$0.15 &1.24 (296)\\
3&0.95  &2.95$\pm$0.07 &6.37$\pm$0.17 &1.23 (294)\\
4&0.97  &2.82$\pm$0.08 &5.53$\pm$0.16 &1.21 (264)\\
\hline
\end{tabular}
\\
\tablefoot{The normalisation factor of the $apec$ model (with abundances set to solar) is related to the EM following $norm=10^{-14} \int n_e n_{\rm H} dV/4\pi d^2=10^{-14} EM/4\pi d^2$. Errors (found using the ``error'' command for the spectral parameters) correspond to 1$\sigma$; whenever errors are asymmetric, 
the largest value is provided here.}
\end{table*}

\begin{table*}
\centering
\caption{Results of the spectral fits of \xmm\ observations over the whole
energy range (0.3--10.0\,keV). The first part of the Table presents
fits with a common absorbing column in front of the soft and
hard components, whereas the second part shows results for separate
absorbing column densities.}
\label{eg2spectralfits}
\setlength{\tabcolsep}{3.3pt}
\begin{tabular}{lccccccccccc}
\hline\hline
\multicolumn{6}{l}{Model $wabs_{ism}*phabs*(apec+apec)$}\\
ID  & $\phi$ & $N_{\rm H}$ & $kT_1$ & $norm_1$ & $kT_2$ & $norm_2$ & $\chi^2$ (dof) & $F^{\rm obs}_{\rm X}$ & 
$F^{\rm unabs}_{\rm X}$\\
&& (10$^{22}$\,cm$^{-2}$) & (keV) & ($10^{-3}$\,cm$^{-5}$) & (keV) & ($10^{-3}$\,cm$^{-5}$) & & \multicolumn{2}{c}{($10^{-12}$\,erg\,cm$^{-2}$\,s$^{-1}$)}\\
\hline
1& 0.57  &0.43$\pm$0.04 & 0.76$\pm$0.02 & 2.55$\pm$0.23 & 3.01$\pm$0.09 & 2.92$\pm$0.09 & 1.18 (492) & 2.56$\pm$0.03 & 5.26 \\
2& 0.89  &0.45$\pm$0.03 & 0.78$\pm$0.02 & 3.79$\pm$0.28 & 3.01$\pm$0.05 & 5.75$\pm$0.10 & 1.23 (575) & 4.79$\pm$0.04 & 9.13 \\
3& 0.95  &0.89$\pm$0.03 & 0.84$\pm$0.03 & 4.35$\pm$0.32 & 3.07$\pm$0.06 & 6.41$\pm$0.15 & 1.29 (561) & 5.00$\pm$0.04 & 7.71 \\
4& 0.97  &1.15$\pm$0.05 & 0.86$\pm$0.04 & 3.32$\pm$0.29 & 3.08$\pm$0.09 & 5.29$\pm$0.19 & 1.52 (511) & 3.91$\pm$0.04 & 5.56 \\
\hline
\multicolumn{6}{l}{Model $wabs_{ism}*(phabs*apec+phabs*apec)$}\\
ID  & $N_{\rm H1}$ & $kT_1$ & $norm_1$ & $N_{\rm H2}$ & $kT_2$ & $norm_2$ & $\chi^2$ (dof) & $F^{\rm obs}_{\rm X}$ & 
$F^{\rm unabs}_{\rm X}$ \\
& (10$^{22}$\,cm$^{-2}$) & (keV) & ($10^{-3}$\,cm$^{-5}$) & (10$^{22}$\,cm$^{-2}$) & (keV) & ($10^{-3}$\,cm$^{-5}$) &   
& \multicolumn{2}{c}{($10^{-12}$\,erg\,cm$^{-2}$\,s$^{-1}$)} \\
\hline
1& 0.56$\pm$0.04 & 0.80$\pm$0.04 & 2.51$\pm$0.18 & 0.00$\pm$0.18 & 3.56$\pm$0.13 & 2.44$\pm$0.08 & 1.13 (491) & 2.60$\pm$0.03 & 5.42\\
2& 0.47$\pm$0.03 & 0.78$\pm$0.02 & 3.72$\pm$0.27 & 0.34$\pm$0.06 & 3.10$\pm$0.08 & 5.55$\pm$0.16 & 1.22 (574) & 4.80$\pm$0.04 & 9.12\\
3& 0.77$\pm$0.04 & 0.75$\pm$0.02 & 4.88$\pm$0.39 & 1.41$\pm$0.10 & 2.67$\pm$0.14 & 7.73$\pm$0.45 & 1.22 (560) & 4.95$\pm$0.04 & 7.86\\
4& 0.90$\pm$0.05 & 0.72$\pm$0.03 & 4.55$\pm$0.39 & 2.69$\pm$0.20 & 2.34$\pm$0.06 & 8.14$\pm$0.33 & 1.21 (510) & 3.87$\pm$0.04 & 5.82\\
\hline
\end{tabular}
\\
\tablefoot{In all cases the interstellar column ($wabs_{ism}$) was fixed to 1.$\times$10$^{22}$\,cm$^{-2}$ (see Sect.~3) and abundances set to solar. ``Unabsorbed'' fluxes are corrected for the interstellar column only. Errors (found using the ``error'' command for the spectral parameters and the ``flux err'' command for the fluxes) correspond to 1$\sigma$; whenever errors are asymmetric, the largest value is provided here. Fluxes are expressed in the 0.5--10.0\,keV band. The $N_{\rm H2}$ value for the XMM-1 fit is basically unconstrained.}
\end{table*}

\subsubsection{Look at the \xmm\ spectra}
The four \xmm\ pn spectra are shown in Fig.~\ref{spec} 
to facilitate the inspection of the changes. With each pn spectrum, 
the best-fit model of the 
pn spectrum preceding it in phase is also shown (see Sect.~4.2.2). 
A simple visual inspection of the spectra confirms the changes 
seen in the light curves. When going from the first to the 
second spectrum, it is evident
that the star brightens. The increase occurs in a very similar way 
(in log flux scale) over the full energy range. 
Going from the second to the third spectrum, the star overall becomes fainter, but this time there are changes with energy; the hard tail (above 3.0\,keV) still exhibits a small increase in flux whilst a strong decrease is seen at lower energies (below 2.0\,keV), suggesting an effect of absorption. As the star becomes even fainter (fourth observation), even the hard part of the spectrum starts to decline. Going from the
fourth observation to the first, the star is still faint in the hard band whereas it is brighter below 1.5\,keV.
Looking at {\it{Swift}} and {\it{Chandra}} spectra confirms these trends, but also reveals the absence of large spectral changes at phases $\phi=0.2-0.8$ as well as the progressive recovery of the soft flux at $\phi=0.0-0.1$. To pinpoint these changes, the spectra were fitted within Xspec v12.8.2 \citep{dor01} assuming solar abundances of \citet{aspl09} and cross-sections of \citet[with changes from \citealt{bal98}, i.e. $bcmc$ case within Xspec]{bal92}. 

\subsubsection{Analysis of the \xmm\ spectra}
In general, we studied the MOS1, MOS2, and pn spectra as well as 
their combination over the three detectors. For the sake of simplicity,
in the following, we limit the description of our results to the combined datasets. 
As a first step, we studied the spectra
restricted to above 3.0\,keV. 
In this region, only a very huge absorbing column
could produce some effect. This provides the opportunity for a detailed
study of the hard band. 
We considered a {\tt{powerlaw}} model, and a {\tt{gaussian}} function for the Fe-K line. The results are given in the top of Table~\ref{egspectralfits}. The simultaneous fits to pn and MOS spectra with this model indicate no significant change in slope, which is entirely compatible with the 1.8 value reported by
\citet{reig99}. Moreover, the position of the Fe-K line is derived
to be in the range 6.67-6.72\,keV. The transitions corresponding to
weakly ionised iron should rather be situated around 6.4\,keV. 
A line located above 6.6\,keV implies that the ion should at least
be \ion{Fe}{xxiii}. The most likely dominant origin is
the \ion{Fe}{xxv} ion as suggested in \citet{ramo16}.  
This indicates that the plasma is highly ionised, as expected 
from strong shocks in a CWR. As the presence of this line indicates 
that the hard X-rays originate in a very hot plasma rather than 
through non-thermal processes, we also fitted a 
mono-temperature unabsorbed {\it{apec}} model 
(see bottom part of Table~\ref{egspectralfits}).
Although the derived temperature gradually shifts from 3.2\,keV to
2.8\,keV when going from spectra XMM-1 to XMM-4, the 
change is not significant (difference less than 3 individual $\sigma$) 
and we can conclude that the hot component does not strongly 
vary in temperature. Therefore, the marked  
variability of the hard flux
is not due to changes in the shock temperature.

\begin{figure*}
\includegraphics[width=8.5cm]{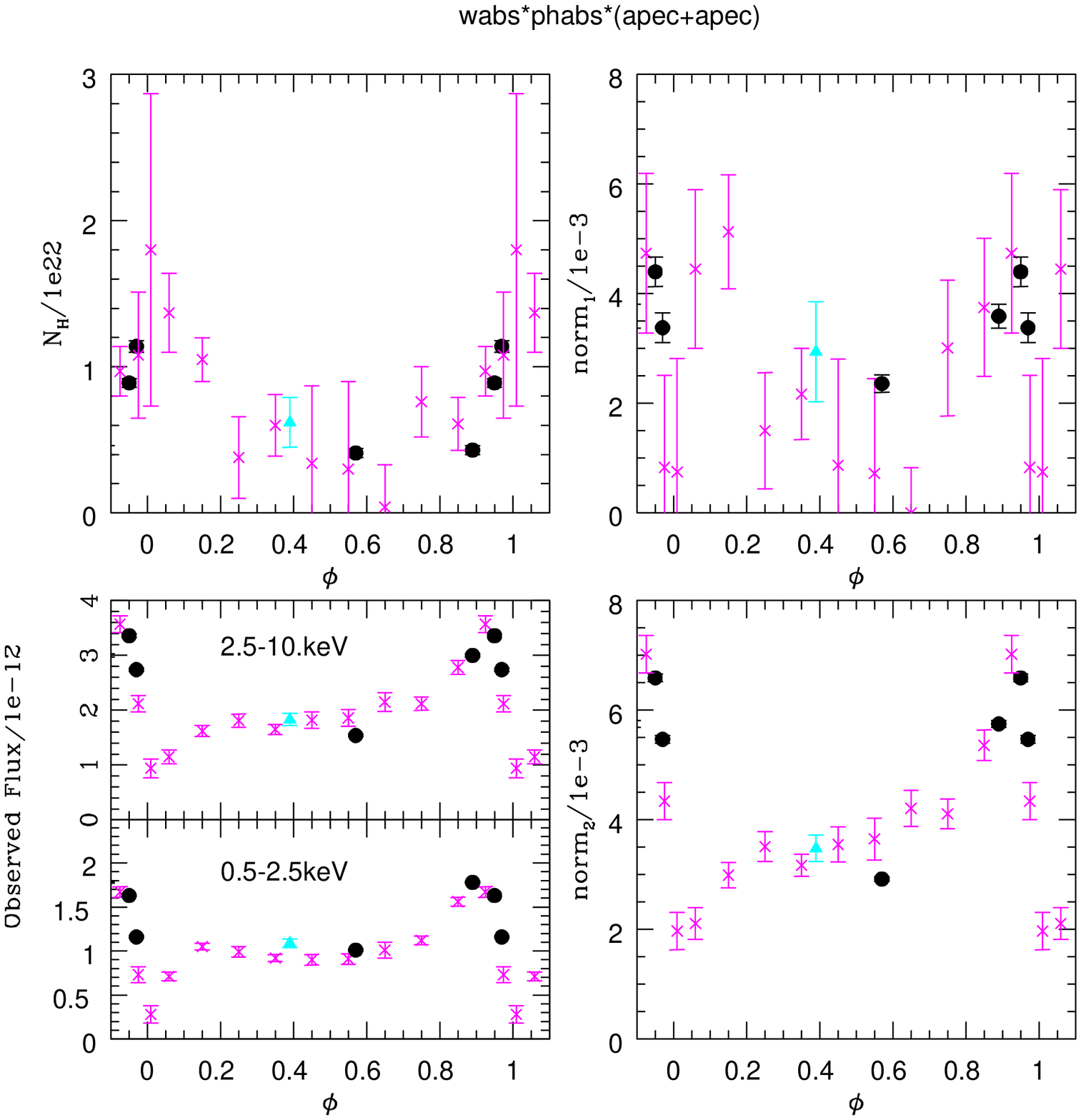}
\hfill
\includegraphics[width=8.5cm]{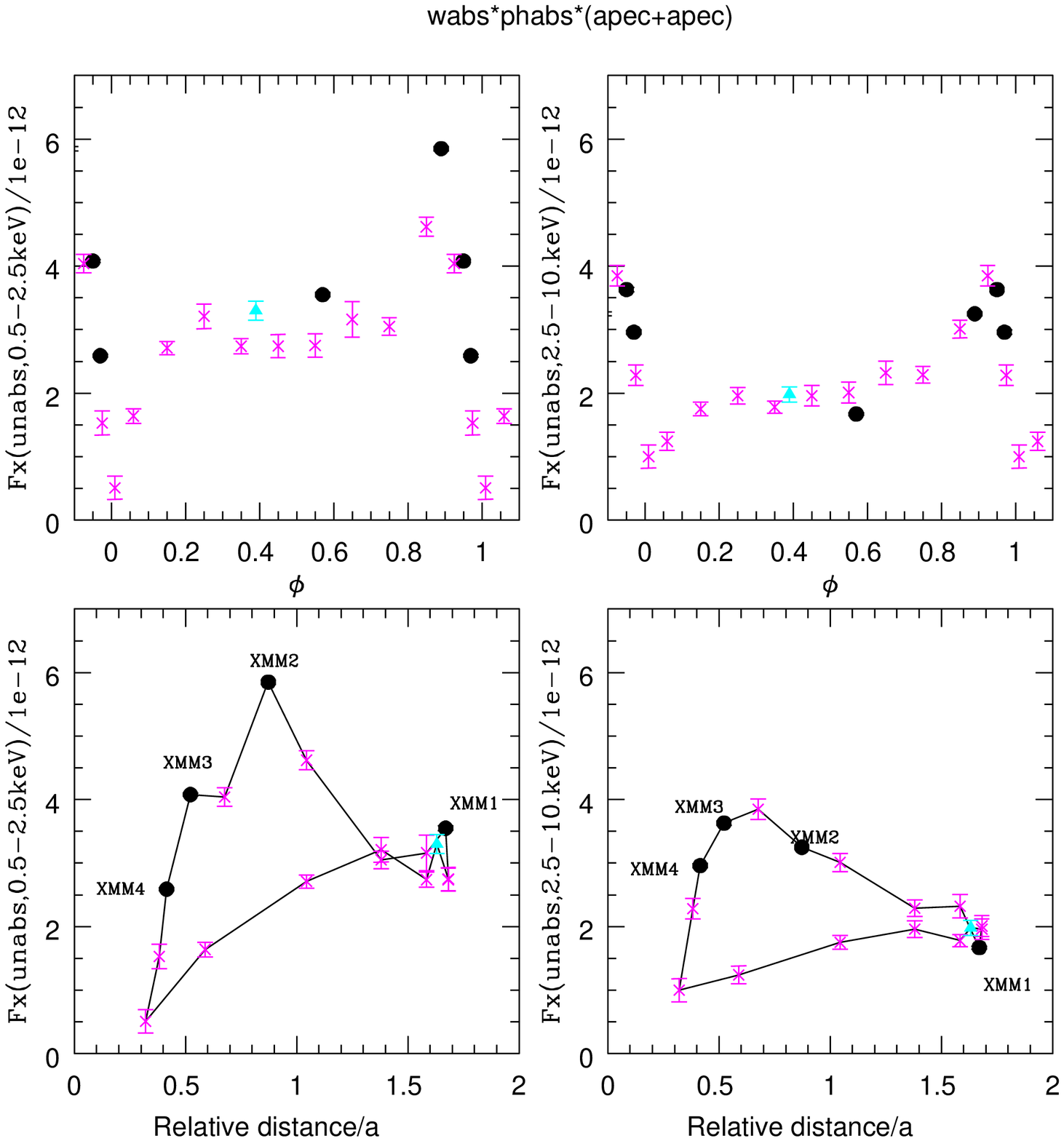}
\caption{Variation with orbital phase of the spectral fit parameters and fluxes (see Table~\ref{spectralfits} for details). Black dots correspond to \xmm\ spectra, the cyan filled triangle to the {\it{Chandra}} spectrum, and magenta crosses to binned {\it{Swift}} spectra (see text). {\it Left panel:} absorption, normalisation factors, and observed fluxes. {\it Right panel:} ISM-absorption corrected fluxes, in both the 0.5--2.5\,keV and 2.5--10.0\,keV energy bands, as a function of phase (top) and separation (bottom).}
\label{specfit}
\end{figure*}

The \xmm\ spectra were also fitted over the whole
energy range (0.3-10.0\,keV) with two-temperature thermal plasma 
models that are optically thin. Again, all three \xmm -EPIC spectra 
of a single observation were simultaneously fitted. 
We first considered a common absorbing
column in front of the two thermal components. Alternatively, 
we considered a model with separate absorbing components in front of the
emitting components. The results are presented in
Table~\ref{eg2spectralfits}. In the former case (common absorbing column), 
the warm component has a temperature of $\sim$3.0\,keV whereas 
in the latter case (individual columns), the fits are
less secure and the warm temperature is around 2.3 to 3.6\,keV, but both
values are in good general agreement with previous results. Concerning the low-temperature component, the temperature is also stable 
and near 0.8\,keV.
The fits with individual absorptions appear better, although the 
difference is only significant for pointing XMM-4.
The absorbing column density in the common absorption fits seems to 
vary from 0.4 to 1.2$\times$10$^{22}$ cm$^{-2}$ but the increase 
is less marked for the absorbing column in front of the soft 
emitter in the two-column fits (0.4 to 0.9$\times$10$^{22}$ cm$^{-2}$).
In these two-column fits, the absorbing column in front of 
the hard component is not very well defined,
as could be expected in view of the lower sensitivity to this parameter, 
but it appears to reach higher values than those in front of the soft component 
(up to $\sim$3$\times$10$^{22}$ cm$^{-2}$). 
Clearly, the strongest variations with phase appear in the intrinsic 
strength of the components, as traced by the EM (or, equivalently, 
the normalisation factors of the $apec$ models). 
From phase 0.57 to 0.89, the EM increases
by 50\% for the soft component and by 100\% for the hard component.
From phase 0.89 to 0.95, the EMs are still increasing
by a few tens of percent. Finally, from phase 0.95 to 0.97, the EMs
for the soft component are decreasing, but the behaviour 
associated with the hard one is less marked.
The decrease in hard flux between XMM-3 and XMM-4 can be 
interpreted as due to a variation of the absorbing
column by an amount of a few 10$^{22}$~cm$^{-2}$.
It should, however, be noted that the decrease in the hard flux 
between the pointings
XMM-3/XMM-4 and the {\it{Swift}} observation at minimum 
(00032960018, see Table~\ref{journal}; 
also labelled Swift-A, see Table~\ref{spectralfits}) is much larger.
It can be reproduced by fixing the norm factor and increasing the
attenuation in front of the hot component by some
3.-4.$\times 10^{23}$~cm$^{-2}$. This is not however the
natural best-fit situation (which favours a decrease 
of the EMs; see below Table~\ref{spectralfits}). 
In addition, such a large column is not
very likely; at least, it should induce
a slope effect inside the 2.5-7.0\,keV range. This effect is not
seen on the Swift-A spectrum. However, we must admit that 
the very low quality of
these {\it{Swift}} data does not
provide a strong constraint. A firmer conclusion would be
obtained by the future acquisition of an \xmm\ observation
at periastron. 
Therefore, we conclude with some caution that a column
increase alone could not be invoked to explain the decrease in flux from
$\phi$\,=\,0.95 to $\phi\sim$ 0.0, and that a decrease of the
intrinsic strength is necessary (i.e.\ a change in the quantity
of emitting material as traced by the emission measure EM).
We thus suggest, in good agreement with the various fits, that the dip
in the X-ray light curve around periastron is due to a combination
of an increase of the absorbing column density and of a decrease
of the EM.  

Finally, we should mention that freeing abundances of HeCNO 
(as the WR component may have non-solar abundances) for either 
the emission or the absorption component neither improves 
the fits nor changes the trends significantly, hence, we restrict 
the discussion to the solar case. 

\begin{sidewaystable*}
\centering
\caption{Results of general spectral fits to \xmm , {\it{Swift}}, and 
{\it{Chandra}} data over the whole energy range and with models
similar to those of Table~\ref{eg2spectralfits} but with fixed temperatures.}
\label{spectralfits}
\begin{tabular}{lcccccccccc}
\hline\hline
\multicolumn{6}{l}{Model $wabs_{ism}*phabs*(apec+apec)$}\\
ID & $\phi$ & $N_{\rm H}$ & $norm_1$ & $norm_2$ & $\chi^2$ (dof) & \multicolumn{2}{c}{$F^{\rm obs}_{\rm X}$} & \multicolumn{2}{c}{$F^{\rm unabs}_{\rm X}$}\\
 & & ($10^{22}$\,cm$^{-2}$) & \multicolumn{2}{c}{($10^{-3}$\,cm$^{-5}$)} & & {\tiny 0.5--2.5\,keV} & {\tiny 2.5--10.0\,keV} & {\tiny 0.5--2.5\,keV} & {\tiny 2.5--10.0\,keV} \\ 
&&&&&&\multicolumn{2}{c}{($10^{-12}$\,erg\,cm$^{-2}$\,s$^{-1}$)}&\multicolumn{2}{c}{($10^{-12}$\,erg\,cm$^{-2}$\,s$^{-1}$)}\\
\hline
1   &0.57  &0.41$\pm$0.03 &2.36$\pm$0.16 &2.92$\pm$0.04 &1.17 (495) &1.01$\pm$0.01 &1.54$\pm$0.02 &3.55 &1.67 \\
2   &0.89  &0.43$\pm$0.03 &3.59$\pm$0.22 &5.75$\pm$0.06 &1.22 (579) &1.78$\pm$0.01 &3.00$\pm$0.03 &5.85 &3.25 \\
3   &0.95  &0.89$\pm$0.03 &4.40$\pm$0.27 &6.59$\pm$0.07 &1.29 (565) &1.63$\pm$0.01 &3.36$\pm$0.03 &4.08 &3.63 \\
4   &0.97  &1.14$\pm$0.04 &3.38$\pm$0.27 &5.47$\pm$0.07 &1.53 (514) &1.16$\pm$0.01 &2.74$\pm$0.03 &2.59 &2.96 \\
A   &0.01  &1.80$\pm$1.07 &0.75$\pm$2.07 &1.97$\pm$0.34 &0.89 ~(10)  &0.28$\pm$0.10 &0.94$\pm$0.17 &0.51 &1.00 \\
B   &0.06  &1.37$\pm$0.27 &4.45$\pm$1.45 &2.11$\pm$0.29 &1.39 ~(34)  &0.71$\pm$0.05 &1.15$\pm$0.13 &1.64 &1.24 \\
C   &0.15  &1.05$\pm$0.15 &5.13$\pm$1.04 &2.99$\pm$0.23 &0.96 (113) &1.05$\pm$0.04 &1.62$\pm$0.10 &2.71 &1.75 \\
D   &0.25  &0.38$\pm$0.28 &1.50$\pm$1.06 &3.51$\pm$0.27 &0.71 ~(76)  &0.99$\pm$0.06 &1.81$\pm$0.12 &3.21 &1.96 \\
E   &0.35  &0.60$\pm$0.21 &2.17$\pm$0.83 &3.17$\pm$0.20 &0.97 (132) &0.92$\pm$0.04 &1.65$\pm$0.09 &2.74 &1.78 \\
F   &0.45  &0.34$\pm$0.53 &0.87$\pm$1.94 &3.55$\pm$0.32 &0.97 ~(58)  &0.90$\pm$0.06 &1.82$\pm$0.15 &2.74 &1.96 \\
G   &0.55  &0.30$\pm$0.60 &0.72$\pm$1.73 &3.65$\pm$0.38 &1.23 ~(41)  &0.91$\pm$0.06 &1.86$\pm$0.15 &2.75 &2.01 \\
H   &0.65  &0.04$\pm$0.29 &0.00$\pm$0.83 &4.21$\pm$0.33 &0.60 ~(46)  &1.01$\pm$0.09 &2.15$\pm$0.17 &3.16 &2.32 \\
I   &0.75  &0.76$\pm$0.24 &3.01$\pm$1.24 &4.11$\pm$0.27 &0.99 (144) &1.12$\pm$0.05 &2.12$\pm$0.12 &3.05 &2.29 \\
J   &0.85  &0.61$\pm$0.18 &3.75$\pm$1.26 &5.36$\pm$0.28 &1.08 (202) &1.56$\pm$0.05 &2.78$\pm$0.13 &4.62 &3.01 \\
K   &0.925 &0.97$\pm$0.17 &4.74$\pm$1.46 &7.02$\pm$0.34 &1.04 (159) &1.67$\pm$0.06 &3.57$\pm$0.15 &4.04 &3.85 \\
L   &0.975 &1.08$\pm$0.43 &0.83$\pm$1.68 &4.34$\pm$0.34 &1.78 ~(54)  &0.73$\pm$0.09 &2.12$\pm$0.15 &1.53 &2.28 \\
203 &0.39  &0.62$\pm$0.17 &2.94$\pm$0.91 &3.48$\pm$0.24 &0.81 (77)  &1.09$\pm$0.05 &1.83$\pm$0.11 &3.30 &1.98 \\
\hline
\multicolumn{6}{l}{Model $wabs_{ism}*(phabs*apec+phabs*apec)$}\\
ID & $\phi$ & $N_{\rm H1}$ & $norm_1$ & $N_{\rm H2}$ & $norm_2$ & $\chi^2$ (dof) & \multicolumn{2}{c}{$F^{\rm obs}_{\rm X}$} & \multicolumn{2}{c}{$F^{\rm unabs}_{\rm X}$} \\
 & & ($10^{22}$\,cm$^{-2}$) & ($10^{-3}$\,cm$^{-5}$) & ($10^{22}$\,cm$^{-2}$) & ($10^{-3}$\,cm$^{-5}$) & & {\tiny 0.5--2.5\,keV} & {\tiny 2.5--10.0\,keV} & {\tiny 0.5--2.5\,keV} & {\tiny 2.5--10.0\,keV} \\ 
&&&&&&&\multicolumn{2}{c}{($10^{-12}$\,erg\,cm$^{-2}$\,s$^{-1}$)}&\multicolumn{2}{c}{($10^{-12}$\,erg\,cm$^{-2}$\,s$^{-1}$)}\\
\hline
1  & 0.57 & 0.44$\pm$0.03 & 2.06$\pm$0.16 & 0.17$\pm$0.07 & 2.86$\pm$0.05 & 1.15(494) & 1.01$\pm$0.01 & 1.52$\pm$0.02 & 3.60 & 1.65\\
2  & 0.89 & 0.44$\pm$0.03 & 3.48$\pm$0.23 & 0.38$\pm$0.05 & 5.72$\pm$0.06 & 1.22(578) & 1.78$\pm$0.01 & 2.99$\pm$0.03 & 5.84 & 3.23\\
3  & 0.95 & 0.83$\pm$0.03 & 4.99$\pm$0.33 & 1.20$\pm$0.08 & 6.77$\pm$0.08 & 1.25(564) & 1.62$\pm$0.01 & 3.41$\pm$0.04 & 4.18 & 3.68\\
4  & 0.97 & 0.98$\pm$0.05 & 4.33$\pm$0.35 & 1.93$\pm$0.13 & 5.85$\pm$0.08 & 1.35(513) & 1.15$\pm$0.01 & 2.85$\pm$0.03 & 2.77 & 3.06\\
203& 0.39 & 0.99$\pm$0.66 & 2.70$\pm$0.88 & 0.16$\pm$0.29 & 3.30$\pm$0.28 & 0.79~(76) & 1.10$\pm$0.07 & 1.77$\pm$0.11 & 3.26 & 1.91\\
\hline
\end{tabular}
\\
\tablefoot{In all cases the interstellar column ($wabs_{ism}$) was fixed to 1.$\times$10$^{22}$\,cm$^{-2}$ (see Sect.~3), temperatures to 0.8 and 3.0\,keV, and abundances to solar. ``Unabsorbed'' fluxes are corrected for the interstellar column only. Errors (found using the ``error'' command for the spectral parameters and the ``flux err'' command for the fluxes) correspond to 1$\sigma$; whenever the errors are asymmetric, the largest value is provided here. Items A--L correspond to a simultaneous fit of {\it{Swift}} spectra in 0.0--0.05, 0.05--0.1, 0.1--0.2, 0.2--0.3, 0.3--0.4, 0.4--0.5, 0.5--0.6, 0.6--0.7, 0.7--0.8, 0.8--0.9, 0.9--0.95, 0.95--1.0 phase bins, respectively (see Table~\ref{journal}, the A and B bins contain only one spectrum each, observations 00032960018 and 00032960002, respectively).}
\end{sidewaystable*}

\subsection{Analysis of the whole dataset}
All the X-ray spectra were fitted within a general scheme 
comprising a common model. Since the signal to noise of the individual 
{\it{Swift}} spectra are very low (about 80--150 raw counts), 
we defined 12 phase bins (0.0--0.05, 0.05--0.1, 0.1--0.2, 0.2--0.3, 0.3--0.4, 0.4--0.5, 0.5--0.6, 0.6--0.7, 0.7--0.8, 0.8--0.9, 0.9--0.95, 
and 0.95--1.0) and simultaneously fitted the spectra taken within 
the same phase bin. For fitting all these spectra, two thermal 
components are sufficient to provide a good fit, as shown before for XMM data.
Temperatures of about 0.8 and 3.0\,keV were always found, hence, 
we decided to fix the temperatures to these values. 
However, it must be noted that models with individual absorptions 
proove too erratic when fitted on the noisier {\it{Swift}} spectra, so 
we restricted ourselves to the common absorption model for 
these data (it clarifies the trends). Fitting results are 
provided in Table~\ref{spectralfits} and shown graphically on Fig.~\ref{specfit}. There are some small differences between 
{\it{Swift}}, {\it{Chandra}}, and \xmm\ results, probably because 
of noise and remaining cross-calibration effects, but these 
differences remain well within errors.

Fluxes and normalisation factors mirror the trends seen in the 
count rates with an increase before periastron, then a sharp drop, 
and finally a slow recovery (Fig.~\ref{specfit}). It should be 
particularly underlined that these variations are not restricted to 
the cooler emission component or the softest flux, despite the 
fact that hard X-rays are much less influenced by absorption effects. 
In addition, looking at normalisation factors, it is also clear that 
the maximum intrinsic emission occurs near the third \xmm\ observation
($\phi\sim 0.95$). If the maximum observed soft flux (or count rate) 
occurs in the second observation ($\phi\sim 0.89$), this is solely due to the 
increased local absorption, which significantly decreases the soft 
flux (indeed, the maximum {\it hard} flux occurs near $\phi\sim 0.95$). 

Local absorption indeed varies, especially at phases 0.9--1.1. The 
maximum is recorded at $\phi\sim0.0$ with at least a tripling of 
the value corresponding to the passage of the WR star in front of 
the system (Table~\ref{spectralfits} and Fig.~\ref{specfit}). When 
two individual absorptions are used (see bottom of 
Table~\ref{spectralfits}), they are similar at first, but that of 
the hotter component increases more rapidly towards periastron 
than that of the cooler component. This agrees well with CWR models 
\citep{pipa10}, which predict the generation of hard X-ray flux 
deeper in the winds, near the line of centres (whilst soft emission 
may arise further downwards of the shock cone). 

It must be noted that these variations are not symmetrical around 
periastron. First, the decline in intrinsic emission begins before 
periastron. Second, the increase in absorption occurs more rapidly 
than the decrease. This leads to an interesting hysteresis effect 
(see right panels of Fig.~\ref{specfit}), which is very similar to 
predictions of \citet[][see next section for a discussion]{pipa10}.      

\section{Discussion}
The X-ray monitoring of WR\,21a brought several important results.
We now summarise and tentatively interpret them, also
pointing out the remaining open questions. 
Although the orbit of WR\,21a is rather well determined, this is far from
the case for the other properties of the two components 
(mass-loss rates, radii, ....) as already mentioned in Sect.~3.2. 
In this context, the discussion could only be considered as preliminary. 
We divided the discussion into four basic topics.
\subsection{The X-ray light curve from phase 0.2 to 0.9}
As mentioned in previous sections, the X-ray light curves
seem to exhibit a $1/D$ flux variation (where $D$ is the 
instantaneous separation between both stars). 
The similarity of the observed variations with a $1/D$ trend 
is so high that it is unlikely that it could be a chance effect.
The separation at $\phi$~=~0.5 is 1.694$\times a$  
(with $a$ the semimajor axis of the orbit) and that 
at $\phi$~=~0.9 is 0.812$\times a$. 
The ratio between separations amounts to 2.08, which 
agrees well with the above-mentioned 100\% increase 
in fluxes and count rates between these phases.
Such a $1/D$ behaviour is expected when cooling of the 
post-shock gas is adiabatic \citep{stev92}. 
The effect arises because the emission volume 
scales as $D^3$ whilst the emissivity per unit volume goes 
as the square of the density and this density
(pre-shock and post-shock) scales as $\dot{M}/D^2$, as explained by
\citet[][see their sect.~3.1 and Eq.10]{stev92}.
Such a $1/D$ trend was detected in some long-term WR+O systems
like WR\,140 \citep[period of 7.9 years; ][]{corc11} 
and WR\,25 \citep[period of 208\,d; ][]{goss07, guna10, pand14} but 
it was not observed in the complex case of
WR\,22 \citep[period of 80\,d; ][]{goss09, park11}.
 
However, based on the probable stellar and wind 
parameters (Table~\ref{orb}), the shocked WR wind in 
WR\,21a should instead 
be on the rapidly cooling side (see Sect.~3.2) and this WR wind 
should dominate the emission (although this is 
still to be proven). 
The $1/D$ behaviour is therefore somewhat unexpected.  
Nevertheless, this discrepancy could perhaps be 
understood considering the work
of \citet{zhek12}, which seemed to detect an adiabatic behaviour
in a sample of close WR+O binaries.
If plotted in figure 2 (right panel) of
\citet{zhek12},  WR\,21a would be precisely located on the
solid straight line expected for systems with a CWR in the
adiabatic regime.
\citet{zhek12} concludes that the
phenomenon making short-period systems exhibit an adiabatic 
behaviour could be due to the 
clumpy nature of the wind. 

An alternative possibility could perhaps be the fact that the companion 
wind is adiabatic. As shown in \citet{park11}, in the case
of the preliminary models of WR\,22, the thermal pressure of the O wind
in the post-shock region acts like a cushion for the contact
discontinuity. This has two effects. First, the contact discontinuity
does not change its general structure. Second, this cushion prevents
the thin-shell instabilities (generated by e.g. the radiative cooling 
of the WR wind) from growing in a non-linear manner. In such a case, one
could imagine that instabilities in the WR wind are damped and, hence, 
not strong enough
to destroy the above-mentioned density effect leading to the $1/D$
behaviour. Only detailed 3D hydrodynamical simulations could perhaps
elucidate this surprising observation but such simulations 
are well beyond the scope of the present paper.
\subsection{The X-ray light curve around periastron}
In the regions of the light curve where the stars are approaching 
periastron, the flux of WR\,21a exhibits a sudden 
decrease and a much slower recovery. 
Such a behaviour has been observed in WR\,140 \citep{corc11} and in $\eta$ Carinae \citep[][and references therein]{moco09}. 
Is the present one of the same nature\,?
The minimum flux appears near $\phi\sim\,$0.9935 in the soft band. This 
phase corresponds to conjunction (with the WR star in front)
and thus to a possible eclipse.
The typical width at half depth of the observed dip is 0.07 (in phase).
A core-core eclipse would span a time corresponding to
a maximum of $\delta \phi$~=~0.01-0.015. 
In order to generate the wider dip that is observed, 
the CWR would have to be
more extended than the O stellar core, which is certainly possible.
However, only part of the CWR would then be eclipsed  
by the small WR core at a given phase, failing to reproduce the
strong decrease in hard flux.  
In addition, the absence of eclipses is noted in the UV light curve
(see Fig.~\ref{uvot}), yielding support for a low inclination angle.
Moreover, the favoured inclination derived from the orbital solution 
and typical masses of O-stars (58\fdg 8) would also prevent eclipses 
from occurring. Therefore, the decrease of the X-ray fluxes should 
be linked to a phenomenon with a longer lasting effect, such as 
an increase of absorption along the line of sight due to the 
dense WR wind, as also suggested by Fig.~\ref{spectralfits}. 
However, we should note that the WR is the star closer to the 
observer between $\phi=0.928$ and 0.0335, whilst
the minimum flux episode occurs from $\phi=0.95$ to 0.15. 
This is a large difference; a
Coriolis effect is probably
not able to fully explain it, but detailed simulations need 
to be performed to pinpoint the extent of the problem.

The corresponding minimum in the hard band occurs 
at $\phi$~=~0.0. As explained in Sect.~4, it is more difficult
to explain by absorption since it would necessitate an
extremely large column of 3.-4.$\times 10^{23}$~cm$^{-2}$
as a result of the much lower sensitivity of the hard X-rays to
photo-absorption. Such a huge column density is unlikely even if 
it is now generally admitted that the structure of WR winds 
very close to their hydrostatic surface is not properly described by 
the current models.
Therefore, another cause is responsible for the hard 
X-ray flux decrease.
The minimum in the hard band essentially
corresponds to a decrease of the EMs. 
An apparent decrease of the EM could be obtained through
a core-core eclipse phenomenon but, as already stated above,
this hypothesis is unlikely.  
On the other hand, the stability of the colliding zone may be questioned:
When stars are close to each other, a stable CWR may not be reached, 
and the stronger wind may then push the CWR to the vicinity of the O-star;
this leaves less room for the O-star wind to accelerate. 
This would lead to a crash of the WR wind onto (or close to) 
the photosphere of the companion. The slow recovery after 
a disruption could further explain the slow rise 
of the fluxes after periastron 
(up to phase $\phi\sim$\,0.1-0.2 -- for an illustration; see Model B of \citealt{park11}). 
The uncertainty on the efficiency of radiative braking 
does not permit us to be predictive without a detailed modelling
but, using the parameters from
Table~\ref{orb}, the WR wind could overwhelm the O wind starting
around phase 0.8-0.9. 
However, the disruption of the CWR should lead 
to the disappearance of the hard flux, as predicted by hydrodynamic 
simulations \citep{park11}. Here, on the contrary, WR\,21a just becomes 
fainter;  its hard X-ray flux never completely disappears.

Figure~\ref{specfit} shows that the variation of the fluxes
with stellar separation presents a strong hysteresis. 
Such a phenomenon has been observed in several systems 
\citep[see e.g. the case of CygOB2\#8A in][]{cazo14}. 
Theoretically, this effect has been detected by 
\citet{pipa10} when analysing
the light curve of their model labelled cwb4. 
Considering only the expected changes in luminosity
of a CWR in the adiabatic regime because of the changing
orbital separation and true wind velocity at the shock, 
\citet{suga15} demonstrated that only at most
half of the amplitude of the 
hysteresis phenomenon can be quantitatively
explained; an additional effect is needed to explain its full 
extent, in particular, the region of the central dip. 
Moreover, an apparent change in wind velocity along the orbital 
cycle should induce a change in the shocked plasma temperature. 
This expected change 
is not observed in the data, suggesting that it
is not the solution.
An instability leading to a disruption 
(at least partial) thus seems
necessary to explain the full extent of the hard band variations. 

To explain both the soft and hard band variations
of WR\,21a, both absorption and 
partial disruption are thus necessary. The disruption
is usually marked, however, by a net softening of the X-ray 
emission that is not
present in WR\,21a. All this should, however, be 
further studied and confirmed
using detailed 3D hydrodynamical simulations.
\subsection{The X-ray light curve around $\phi$~=~0.316}
Phase $\phi$~=~0.316 corresponds to the conjunction 
with the O-star in front.
At that particular moment, the cone formed by the CWR is 
mainly directed towards the observer. 
If the density of the O-star wind is lower, which is
almost always the case, then there is less absorption 
along the line of sight and the shock zone is better 
seen through the O-star wind
than during the remaining parts of the orbit. 
This should be accompanied by an apparent brightening of 
the X-ray luminosity. 
However, this effect is only observable if the 
line of sight from the apex
towards the observer lies within the shock cone and 
this is only possible if the complement of the
inclination of the system is less than half the opening angle of 
the CWR. The X-ray light curve of WR\,21a
does not show any variation of this kind at that particular phase,
reinforcing the conclusion that large inclination values 
are excluded for this system.

\subsection{The expected absorbing column density}
\begin{figure}
\includegraphics[width=8.5cm]{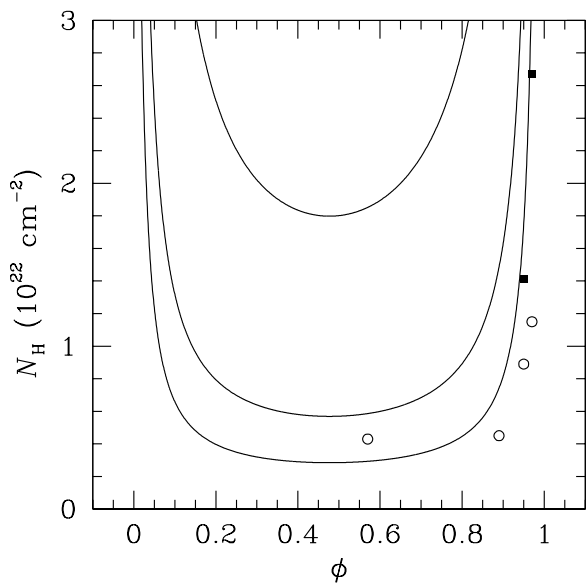}
\caption{Variation with phase of the equivalent hydrogen
column density along the line of sight 
from the apex. The column is computed according to the physical 
parameters given in Table~\ref{orb} and for an inclination of 58\fdg 8. 
We show the curves expected for various mass-loss rates:
0.5$\times$10$^{-5}$ M$_{\sun}$\,yr$^{-1}$ (lower curve), 
1.0$\times$10$^{-5}$ M$_{\sun}$\,yr$^{-1}$ (middle curve),
and
3.2$\times$10$^{-5}$ M$_{\sun}$\,yr$^{-1}$ (upper curve, value
adopted in Table~\ref{orb}). The figure also
includes (open circles) the four values for the common column
fits deduced from the \xmm\ observations. In addition,
the two filled squares indicate the column in front of the sole hard component
for the XMM-3 and XMM-4 pointings. 
}
\label{column}
\end{figure}

Just before periastron, the absorbing column along the line of sight
from the apex of the CWR to the observer reaches a maximum
(see e.g. Fig.~\ref{specfit}). 
On the basis of the adopted parameters
given in Table~\ref{orb}, we can calculate a theoretical 
equivalent hydrogen column by integrating the wind densities
found around the WR star (as a function of distance to that star) 
along the line of sight from the expected position of the apex
to a point far away enough from the WR that the wind no longer 
contributes significantly to the column. 
The computed density column is then corrected by 
$X_{\rm{H}}$/$m_{\rm{H}}$, where the numerator is the abundance in mass
of hydrogen and the denominator the mass of the H atom.
We thus obtain an equivalent $N_{\rm{H}}$.    
Figure~\ref{column} shows, for an inclination of 58\fdg 8 
(see Sect.~3.2), 
the evolution of the theoretical column as a function of phase. 
The figure also exhibits the observed values as derived above.
To agree with these observed columns (derived from the \xmm\ data), 
the theoretical mass-loss rate of the WR star would have to be reduced 
to $\sim 1\times10^{-5}$\,M$_{\sun}$\,yr$^{-1}$ at a first level
of approximation.
This result remains true even for non-solar abundances typical of
WNLh stars, as fitting trials with different column models
(solar composition versus that of WR\,22) yielded similar
column values, taking the different $X_{\rm{H}}$ 
into account. However,
no effort has been made to include the extent of the
emitting region in this calculation and
a definitive answer on WR\,21a mass-loss rate should 
await confirmation by sophisticated
hydrodynamical simulations. 
\section{Conclusion}
WR\,21a is a very interesting system as it contains a WNLh star, which 
turns out to be possibly very massive and is a key point to 
understanding massive star evolution.
To learn more about the winds in this system, we have studied 
its wind-wind collision in X-rays using \xmm\ as well as 
archival {\it{Chandra}} and {\it{Swift}} datasets. 

The EPIC spectra are well fitted by a two-component, optically-thin
thermal plasma model suggesting as temperatures 0.8\,keV and 3.0\,keV.
These temperatures are rather well defined and present almost no change. 
The situation appears very different when examining the X-ray fluxes.
From phase $\phi$~=~0.2 to 0.8, the fluxes do not vary much 
but then they increase and at $\phi$~=~0.9 reach a level twice 
that at $\phi$~=~0.5. 
The highest observed flux, in the band 0.5--10.0\,keV, amounts to
$f_{\rm X}^{\rm obs} \, = \,5 \times 10^{-12}$ erg cm$^{-2}$ s$^{-1}$,
and the maximum flux in the same band, corrected for the interstellar absorption, is 
$f_{\rm X}^{\rm unabs} \, = \,9.1 \times 10^{-12}$ erg cm$^{-2}$ s$^{-1}$.
This leads to a luminosity
$L_{\rm X}^{\rm obs} \, = \, 1.1 \times 10^{33}$ $d^2$ erg s$^{-1}$,
if $d$ is the distance to the star given in kpc. 
Assuming a distance of 5.2 kpc, this leads to a
luminosity of $3.0\times 10^{34}$ erg s$^{-1}$,
placing WR\,21a amongst the brightest WN~+~O systems, 
along with e.g. HD\,5980 
\citep[$L_{\rm X}^{\rm obs} \, = \, 1.7 \times 
10^{34}$ erg s$^{-1}$;\,][]{naze02} 
and WR\,25 
\citep[$L_{\rm X}^{\rm obs} \, = \, 1.1 \times 
10^{34}$ erg s$^{-1}$;\,][]{pand14}; 
for comparison, WR\,22 and WR\,20a are one order of magnitude fainter \citep{goss09, naze08}. These conclusions are highly dependent on the
adopted values for the individual distances, however.

From $\phi$~=~0.2 to $\phi$~=~0.9, the X-ray light curve
seems to follow a $1/D$ trend, suggesting the CWR cools 
adiabatically whilst a radiative collision is expected. 
This result is however in line with the conclusions reached 
by \citet{zhek12}. 
After $\phi$~=~0.9, the flux starts to decrease rather rapidly,
reaching a minimum in the soft band at the time of the 
conjunction with the WR star in front.
In the hard band, the minimum occurs slightly later, instead corresponding
to the periastron passage, although the difference in phase of the
two events is very small. The recovery from the minimum
is slower than the decrease and ends at $\phi$~=~0.1-0.2.
Eclipses cannot be considered a potential explanation for this 
variability, as several lines of evidence exclude a high 
inclination value: the duration of the X-ray flux minimum at periastron 
and the absence of increase 
in the X-ray flux at the conjunction with the O-star in front, 
the absence of eclipses in the UV domain, and the values of the stellar masses 
derived from the orbital solution.
The observed decline in flux is then probably due to 
two phenomena. First, there is a 
strong absorption as the WR and its dense wind appear in front. 
This mostly affects the soft band and the X-ray spectral fits 
can then be used to constrain the mass-loss rate of the WR;
we found a preliminary value of 
$\sim 1\times10^{-5}$\,M$_{\sun}$\,yr$^{-1}$. 
Second, the decrease in the intrinsic strength of the X-ray 
emission suggests a (partial) disruption of the shock, 
or even a crash of the CWR onto the photosphere of the 
companion, near or at periastron. After periastron passage, 
the recovery of the emission presents a strong hysteresis effect.

Now that the X-ray variations are well constrained, 
detailed atmosphere analysis of the UV/visible spectra 
and hydrodynamical simulations of the CWR are needed 
to further improve our understanding 
of this extremely massive system.

\begin{acknowledgements}
We acknowledge support from the Fonds National de la Recherche Scientifique (Belgium), the Communaut\'e Fran\c caise de Belgique, the PRODEX \xmm\ and Integral contracts, and the `Action de Recherche Concert\'ee' (CFWB-Acad\'emie Wallonie Europe). We thank Kim Page ({\it{Swift}} UK centre) for her kind assistance. ADS and CDS were used for preparing this document. 
We thanks H.\ Sana for a preliminary reduction of the FEROS data, and for having made his ephemeris available in advance of publication.
\end{acknowledgements}

\newpage
\begin{appendix} 
\section{Journal of the X-ray observations}
\begin{table*}
\centering
\footnotesize
\caption{Journal of the X-ray observations. Mid-exposure phases were calculated using the ephemeris of \citet{tram16}, exposure times correspond to on-axis values (for pn if \xmm). The \xmm\ count rates correspond to the sum of MOS1, MOS2, and pn values.}
\label{journal}
\begin{tabular}{llcccccc}
\hline\hline
XID & Obs. & ObsID (exp. time) & Start Date & HJD & $\phi$ & \multicolumn{2}{c}{Count Rates (ct\,s$^{-1}$)}\\
   &      &                   &            & at mid-exposure 
&        & 0.4--2.0\,keV & 2.0--10.0\,keV \\
\hline
  1$^*$ & XMM & 0724190501 (19.7 ks) & 2013-06-14T20:48:06 & 2456458.513 & 0.57 & 0.672$\pm$0.007 & 0.513$\pm$0.006 \\ 
  2$^*$ & XMM & 0724190601 (19.7 ks) & 2013-07-26T15:34:27 & 2456500.293 & 0.89 & 1.123$\pm$0.009 & 0.969$\pm$0.008 \\ 
  3$^*$ & XMM & 0724190701 (17.9 ks) & 2013-07-28T15:26:08 & 2456502.275 & 0.95 & 0.930$\pm$0.008 & 1.059$\pm$0.009 \\ 
  4$^*$ & XMM & 0724190801 (16.3 ks) & 2013-07-29T03:26:28 & 2456502.847 & 0.97 & 0.570$\pm$0.006 & 0.797$\pm$0.007 \\ 
  5 &\swift & 00032960001 (0.5 ks) & 2013-10-01T20:48:28 & 2456567.371 & 0.01 & 0.009$\pm$0.005 & 0.015$\pm$0.006 \\ 
  6$^*$ &\swift & 00032960002 (9.7 ks) & 2013-10-03T01:34:16 & 2456569.003 & 0.06 & 0.025$\pm$0.002 & 0.025$\pm$0.002 \\ 
  7$^*$ &\swift & 00032960003 (9.7 ks) & 2013-10-05T01:38:11 & 2456570.972 & 0.12 & 0.043$\pm$0.002 & 0.036$\pm$0.002 \\ 
  8$^*$ &\swift & 00032960004 (9.6 ks) & 2013-10-07T01:27:26 & 2456572.860 & 0.18 & 0.035$\pm$0.002 & 0.030$\pm$0.002 \\ 
  9     &\swift & 00032960005 (0.9 ks) & 2013-10-09T00:03:25 & 2456574.505 & 0.23 & 0.042$\pm$0.008 & 0.043$\pm$0.008 \\ 
 10     &\swift & 00032960006 (1.0 ks) & 2013-10-11T20:44:35 & 2456577.367 & 0.32 & 0.045$\pm$0.008 & 0.033$\pm$0.006 \\ 
 11     &\swift & 00032960008 (1.0 ks) & 2013-10-15T09:38:35 & 2456580.904 & 0.43 & 0.035$\pm$0.006 & 0.029$\pm$0.006 \\ 
 12     &\swift & 00032960009 (1.0 ks) & 2013-10-17T19:24:10 & 2456583.311 & 0.51 & 0.034$\pm$0.008 & 0.024$\pm$0.007 \\ 
 13     &\swift & 00032960010 (1.1 ks) & 2013-10-19T14:40:12 & 2456585.114 & 0.57 & 0.043$\pm$0.008 & 0.050$\pm$0.008 \\ 
 14     &\swift & 00032960011 (1.2 ks) & 2013-10-21T02:00:09 & 2456586.683 & 0.62 & 0.031$\pm$0.006 & 0.036$\pm$0.006 \\ 
 15     &\swift & 00032960012 (1.0 ks) & 2013-10-25T16:38:17 & 2456591.196 & 0.76 & 0.036$\pm$0.007 & 0.036$\pm$0.007 \\ 
 16$^-$ &\swift & 00032960013 (1.0 ks) & 2013-10-27T10:21:25 & 2456592.935 & 0.81 & 0.060$\pm$0.008 & 0.035$\pm$0.006 \\ 
 17$^*$ &\swift & 00032960014 (7.9 ks) & 2013-10-29T11:57:02 & 2456595.233 & 0.89 & 0.063$\pm$0.003 & 0.061$\pm$0.003 \\ 
 18$^*$ &\swift & 00032960015 (3.8 ks) & 2013-10-30T08:51:51 & 2456596.103 & 0.91 & 0.060$\pm$0.004 & 0.069$\pm$0.005 \\ 
 19$^*$ &\swift & 00032960016 (8.3 ks) & 2013-10-31T00:43:17 & 2456596.901 & 0.94 & 0.054$\pm$0.003 & 0.070$\pm$0.003 \\ 
 20$^*$ &\swift & 00032960017 (6.9 ks) & 2013-11-01T07:26:54 & 2456598.208 & 0.98 & 0.019$\pm$0.002 & 0.029$\pm$0.002 \\ 
 21$^-$ &\swift & 00032960018 (7.1 ks) & 2013-11-02T08:48:14 & 2456599.132 & 0.01 & 0.009$\pm$0.001 & 0.015$\pm$0.002 \\ 
 22$^*$ &\swift & 00032960019 (5.5 ks) & 2014-06-21T02:28:20 & 2456829.707 & 0.29 & 0.037$\pm$0.003 & 0.036$\pm$0.003 \\ 
 23$^*$ &\swift & 00032960020 (4.9 ks) & 2014-06-22T00:57:10 & 2456830.643 & 0.32 & 0.036$\pm$0.003 & 0.032$\pm$0.003 \\ 
 24$^*$ &\swift & 00032960021 (4.2 ks) & 2014-06-23T15:25:17 & 2456832.313 & 0.37 & 0.036$\pm$0.003 & 0.028$\pm$0.003 \\ 
 25$^*$ &\swift & 00032960022 (5.9 ks) & 2014-06-24T03:56:24 & 2456832.771 & 0.38 & 0.033$\pm$0.003 & 0.037$\pm$0.003 \\ 
 26     &\swift & 00032960024 (1.3 ks) & 2014-10-24T23:31:48 & 2456955.485 & 0.26 & 0.038$\pm$0.006 & 0.027$\pm$0.005 \\ 
 27$^-$ &\swift & 00032960025 (1.3 ks) & 2014-10-25T04:19:47 & 2456955.685 & 0.26 & 0.040$\pm$0.006 & 0.032$\pm$0.005 \\ 
 28     &\swift & 00032960026 (0.9 ks) & 2014-10-25T07:31:23 & 2456955.816 & 0.27 & 0.034$\pm$0.008 & 0.027$\pm$0.007 \\ 
 29$^-$ &\swift & 00032960027 (2.0 ks) & 2014-10-25T10:43:10 & 2456956.015 & 0.27 & 0.035$\pm$0.005 & 0.031$\pm$0.004 \\ 
 30     &\swift & 00032960028 (1.0 ks) & 2014-10-25T18:49:29 & 2456956.287 & 0.28 & 0.033$\pm$0.006 & 0.026$\pm$0.005 \\ 
 31     &\swift & 00032960029 (1.1 ks) & 2014-10-25T23:36:46 & 2456956.487 & 0.29 & 0.038$\pm$0.007 & 0.038$\pm$0.007 \\ 
 32$^-$ &\swift & 00032960030 (1.3 ks) & 2014-10-26T01:07:24 & 2456956.552 & 0.29 & 0.042$\pm$0.006 & 0.032$\pm$0.005 \\ 
 33     &\swift & 00032960031 (1.4 ks) & 2014-10-26T05:52:23 & 2456956.750 & 0.30 & 0.037$\pm$0.009 & 0.025$\pm$0.007 \\ 
 34     &\swift & 00032960032 (1.4 ks) & 2014-10-26T10:40:24 & 2456956.950 & 0.30 & 0.039$\pm$0.006 & 0.025$\pm$0.005 \\ 
 35     &\swift & 00032960033 (0.4 ks) & 2014-10-26T12:38:16 & 2456957.084 & 0.31 & 0.040$\pm$0.010 & 0.041$\pm$0.010 \\ 
 36     &\swift & 00032960034 (1.1 ks) & 2014-10-26T17:05:45 & 2456957.216 & 0.31 & 0.026$\pm$0.006 & 0.037$\pm$0.007 \\ 
 37$^-$ &\swift & 00032960035 (1.4 ks) & 2014-10-26T21:52:45 & 2456957.417 & 0.32 & 0.028$\pm$0.005 & 0.040$\pm$0.006 \\ 
 38     &\swift & 00032960036 (1.3 ks) & 2014-10-27T01:05:12 & 2456957.550 & 0.32 & 0.030$\pm$0.007 & 0.028$\pm$0.006 \\ 
 39$^-$ &\swift & 00032960037 (1.4 ks) & 2014-10-27T05:52:34 & 2456957.750 & 0.33 & 0.026$\pm$0.005 & 0.039$\pm$0.006 \\ 
 40     &\swift & 00032960038 (1.3 ks) & 2014-10-27T09:04:24 & 2456957.883 & 0.33 & 0.035$\pm$0.006 & 0.032$\pm$0.005 \\ 
 41     &\swift & 00032960039 (1.3 ks) & 2014-10-27T13:52:11 & 2456958.112 & 0.34 & 0.032$\pm$0.005 & 0.032$\pm$0.005 \\ 
 42$^-$ &\swift & 00032960040 (1.3 ks) & 2014-10-27T17:04:11 & 2456958.243 & 0.34 & 0.043$\pm$0.006 & 0.031$\pm$0.005 \\ 
 43$^-$ &\swift & 00032960041 (1.4 ks) & 2014-10-27T21:51:24 & 2456958.416 & 0.35 & 0.037$\pm$0.006 & 0.035$\pm$0.005 \\ 
 44     &\swift & 00032960042 (1.4 ks) & 2014-10-28T01:02:26 & 2456958.549 & 0.35 & 0.030$\pm$0.005 & 0.028$\pm$0.005 \\ 
 45$^-$ &\swift & 00032960043 (1.5 ks) & 2014-10-28T04:13:45 & 2456958.682 & 0.36 & 0.036$\pm$0.005 & 0.034$\pm$0.005 \\ 
 46     &\swift & 00032960044 (1.3 ks) & 2014-10-28T09:01:30 & 2456958.881 & 0.36 & 0.032$\pm$0.005 & 0.027$\pm$0.005 \\ 
 47$^-$ &\swift & 00032960045 (1.4 ks) & 2014-10-28T12:13:08 & 2456959.044 & 0.37 & 0.029$\pm$0.005 & 0.034$\pm$0.005 \\ 
 48     &\swift & 00032960046 (1.2 ks) & 2014-10-28T17:02:24 & 2456959.241 & 0.38 & 0.036$\pm$0.006 & 0.023$\pm$0.005 \\ 
 49$^-$ &\swift & 00032960047 (1.5 ks) & 2014-10-28T21:47:44 & 2456959.414 & 0.38 & 0.034$\pm$0.005 & 0.037$\pm$0.005 \\ 
 50     &\swift & 00032960048 (1.4 ks) & 2014-10-29T00:59:45 & 2456959.547 & 0.39 & 0.028$\pm$0.005 & 0.028$\pm$0.005 \\ 
 51$^-$ &\swift & 00032960049 (1.4 ks) & 2014-10-29T05:47:47 & 2456959.747 & 0.39 & 0.036$\pm$0.006 & 0.029$\pm$0.005 \\ 
 52     &\swift & 00032960050 (1.3 ks) & 2014-10-29T10:36:13 & 2456959.946 & 0.40 & 0.041$\pm$0.006 & 0.025$\pm$0.005 \\ 
 53     &\swift & 00032960051 (0.6 ks) & 2014-10-29T12:11:22 & 2456960.009 & 0.40 & 0.040$\pm$0.009 & 0.035$\pm$0.008 \\ 
 54     &\swift & 00032960052 (1.2 ks) & 2014-10-29T17:00:24 & 2456960.212 & 0.41 & 0.025$\pm$0.005 & 0.028$\pm$0.005 \\ 
 55$^-$ &\swift & 00032960053 (1.5 ks) & 2014-10-29T20:10:44 & 2456960.347 & 0.41 & 0.039$\pm$0.006 & 0.022$\pm$0.004 \\ 
 56     &\swift & 00032960054 (1.4 ks) & 2014-10-30T00:58:33 & 2456960.546 & 0.42 & 0.029$\pm$0.005 & 0.027$\pm$0.005 \\ 
 57$^-$ &\swift & 00032960055 (1.4 ks) & 2014-10-30T05:45:46 & 2456960.745 & 0.42 & 0.037$\pm$0.006 & 0.041$\pm$0.006 \\ 
 58$^-$ &\swift & 00032960056 (1.4 ks) & 2014-10-30T08:56:56 & 2456960.878 & 0.43 & 0.036$\pm$0.006 & 0.031$\pm$0.005 \\ 
 59$^-$ &\swift & 00032960057 (1.5 ks) & 2014-10-30T13:44:28 & 2456961.107 & 0.43 & 0.041$\pm$0.006 & 0.035$\pm$0.005 \\ 
 60$^-$ &\swift & 00032960058 (1.5 ks) & 2014-10-30T18:36:33 & 2456961.281 & 0.44 & 0.035$\pm$0.005 & 0.037$\pm$0.005 \\ 
\hline
\end{tabular}
\end{table*}
\setcounter{table}{0}
\begin{table*}
\centering
\footnotesize
\caption{Continued.}
\begin{tabular}{llcccccc}
\hline\hline
ID & Obs. & ObsID (exp. time) & Start Date & HJD & $\phi$ & \multicolumn{2}{c}{Count Rates (ct\,s$^{-1}$)}\\
   &      &                   &            & at mid-exposure &        & 0.4--2.0\,keV & 2.0--10.0\,keV \\
\hline
 61$^-$ &\swift & 00032960059 (1.5 ks) & 2014-10-30T21:48:13 & 2456961.414 & 0.44 & 0.030$\pm$0.005 & 0.034$\pm$0.005 \\ 
 62     &\swift & 00032960060 (1.4 ks) & 2014-10-31T02:32:39 & 2456961.612 & 0.45 & 0.041$\pm$0.009 & 0.044$\pm$0.009 \\ 
 63     &\swift & 00032960061 (1.5 ks) & 2014-10-31T05:43:28 & 2456961.745 & 0.45 & 0.033$\pm$0.008 & 0.036$\pm$0.008 \\ 
 64     &\swift & 00032960063 (0.5 ks) & 2014-10-31T15:18:59 & 2456962.138 & 0.47 & 0.038$\pm$0.009 & 0.032$\pm$0.008 \\ 
 65$^-$ &\swift & 00032960064 (1.5 ks) & 2014-10-31T18:33:59 & 2456962.279 & 0.47 & 0.038$\pm$0.005 & 0.034$\pm$0.005 \\ 
 66$^-$ &\swift & 00032960065 (1.6 ks) & 2014-10-31T21:40:47 & 2456962.410 & 0.48 & 0.038$\pm$0.005 & 0.033$\pm$0.005 \\ 
 67     &\swift & 00032960084 (1.5 ks) & 2014-11-01T18:32:11 & 2456963.278 & 0.50 & 0.032$\pm$0.005 & 0.026$\pm$0.005 \\ 
 68     &\swift & 00032960072 (1.4 ks) & 2014-11-02T02:27:25 & 2456963.608 & 0.51 & 0.030$\pm$0.005 & 0.027$\pm$0.005 \\ 
 69     &\swift & 00032960073 (1.3 ks) & 2014-11-02T07:18:19 & 2456963.809 & 0.52 & 0.030$\pm$0.008 & 0.036$\pm$0.009 \\ 
 70$^-$ &\swift & 00032960074 (1.5 ks) & 2014-11-02T09:07:46 & 2456963.907 & 0.52 & 0.029$\pm$0.005 & 0.034$\pm$0.005 \\ 
 71$^-$ &\swift & 00032960075 (1.5 ks) & 2014-11-02T13:39:26 & 2456964.105 & 0.53 & 0.042$\pm$0.006 & 0.042$\pm$0.006 \\ 
 72     &\swift & 00032960076 (1.4 ks) & 2014-11-02T18:32:46 & 2456964.278 & 0.53 & 0.051$\pm$0.009 & 0.042$\pm$0.009 \\ 
 73$^-$ &\swift & 00032960077 (1.5 ks) & 2014-11-02T21:41:10 & 2456964.409 & 0.54 & 0.036$\pm$0.005 & 0.033$\pm$0.005 \\ 
 74$^-$ &\swift & 00032960078 (1.3 ks) & 2014-11-03T01:03:31 & 2456964.549 & 0.54 & 0.031$\pm$0.005 & 0.038$\pm$0.006 \\ 
 75$^-$ &\swift & 00032960079 (1.5 ks) & 2014-11-03T07:13:25 & 2456964.807 & 0.55 & 0.037$\pm$0.005 & 0.032$\pm$0.005 \\ 
 76     &\swift & 00032960080 (1.2 ks) & 2014-11-03T09:10:42 & 2456964.908 & 0.55 & 0.038$\pm$0.006 & 0.022$\pm$0.005 \\ 
 77     &\swift & 00032960081 (1.3 ks) & 2014-11-03T13:36:31 & 2456965.101 & 0.56 & 0.036$\pm$0.006 & 0.031$\pm$0.006 \\ 
 78     &\swift & 00032960082 (1.4 ks) & 2014-11-03T18:25:41 & 2456965.273 & 0.57 & 0.034$\pm$0.005 & 0.027$\pm$0.005 \\ 
 79     &\swift & 00032960083 (1.4 ks) & 2014-11-03T21:36:40 & 2456965.406 & 0.57 & 0.042$\pm$0.006 & 0.023$\pm$0.004 \\ 
 80     &\swift & 00032960086 (1.6 ks) & 2015-01-06T04:12:07 & 2457028.735 & 0.57 & 0.039$\pm$0.006 & 0.022$\pm$0.004 \\ 
 81     &\swift & 00032960087 (1.4 ks) & 2015-01-06T08:31:14 & 2457028.891 & 0.57 & 0.032$\pm$0.008 & 0.031$\pm$0.007 \\ 
 82$^-$ &\swift & 00032960089 (1.5 ks) & 2015-01-06T16:44:14 & 2457029.206 & 0.58 & 0.032$\pm$0.005 & 0.032$\pm$0.005 \\ 
 83     &\swift & 00032960090 (1.5 ks) & 2015-01-06T21:18:13 & 2457029.396 & 0.59 & 0.037$\pm$0.005 & 0.023$\pm$0.004 \\ 
 84     &\swift & 00032960091 (1.2 ks) & 2015-01-07T00:48:58 & 2457029.541 & 0.59 & 0.038$\pm$0.006 & 0.036$\pm$0.006 \\ 
 85     &\swift & 00032960092 (1.2 ks) & 2015-01-07T02:29:14 & 2457029.663 & 0.60 & 0.041$\pm$0.007 & 0.033$\pm$0.006 \\ 
 86     &\swift & 00032960093 (1.4 ks) & 2015-01-07T08:42:30 & 2457029.931 & 0.61 & 0.029$\pm$0.005 & 0.025$\pm$0.005 \\ 
 87     &\swift & 00032960094 (1.1 ks) & 2015-01-07T13:38:52 & 2457030.075 & 0.61 & 0.046$\pm$0.007 & 0.032$\pm$0.006 \\ 
 88     &\swift & 00032960095 (1.1 ks) & 2015-01-07T16:28:13 & 2457030.255 & 0.62 & 0.028$\pm$0.006 & 0.041$\pm$0.007 \\ 
 89     &\swift & 00032960096 (0.9 ks) & 2015-01-07T21:16:13 & 2457030.392 & 0.62 & 0.044$\pm$0.008 & 0.035$\pm$0.007 \\ 
 90$^-$ &\swift & 00032960097 (1.3 ks) & 2015-01-08T00:45:20 & 2457030.539 & 0.63 & 0.037$\pm$0.006 & 0.033$\pm$0.005 \\ 
 91     &\swift & 00032960098 (1.4 ks) & 2015-01-08T04:07:00 & 2457030.709 & 0.63 & 0.038$\pm$0.007 & 0.027$\pm$0.006 \\ 
 92     &\swift & 00032960099 (1.0 ks) & 2015-01-08T08:27:13 & 2457030.858 & 0.64 & 0.037$\pm$0.007 & 0.025$\pm$0.006 \\ 
 93     &\swift & 00032960100 (0.8 ks) & 2015-01-08T13:40:09 & 2457031.074 & 0.64 & 0.054$\pm$0.011 & 0.045$\pm$0.010 \\ 
 94     &\swift & 00032960101 (0.9 ks) & 2015-01-08T16:50:31 & 2457031.207 & 0.65 & 0.034$\pm$0.007 & 0.031$\pm$0.007 \\ 
 95     &\swift & 00032960102 (1.4 ks) & 2015-01-08T21:33:13 & 2457031.429 & 0.65 & 0.036$\pm$0.006 & 0.033$\pm$0.006 \\ 
 96$^-$ &\swift & 00032960103 (1.4 ks) & 2015-01-09T00:45:00 & 2457031.570 & 0.66 & 0.042$\pm$0.006 & 0.036$\pm$0.005 \\ 
 97$^-$ &\swift & 00032960104 (1.4 ks) & 2015-01-09T04:04:00 & 2457031.706 & 0.66 & 0.038$\pm$0.006 & 0.040$\pm$0.006 \\ 
 98     &\swift & 00032960106 (1.3 ks) & 2015-01-09T13:25:09 & 2457032.068 & 0.67 & 0.037$\pm$0.006 & 0.033$\pm$0.005 \\ 
 99$^-$ &\swift & 00032960107 (1.5 ks) & 2015-01-09T18:14:20 & 2457032.268 & 0.68 & 0.034$\pm$0.005 & 0.037$\pm$0.005 \\ 
100$^-$ &\swift & 00032960108 (1.5 ks) & 2015-01-09T21:25:19 & 2457032.402 & 0.68 & 0.037$\pm$0.005 & 0.042$\pm$0.006 \\ 
101$^-$ &\swift & 00032960109 (1.4 ks) & 2015-01-10T00:41:20 & 2457032.568 & 0.69 & 0.043$\pm$0.006 & 0.045$\pm$0.006 \\ 
102     &\swift & 00032960110 (1.5 ks) & 2015-01-10T04:01:56 & 2457032.739 & 0.70 & 0.040$\pm$0.007 & 0.040$\pm$0.007 \\ 
103$^-$ &\swift & 00032960111 (1.5 ks) & 2015-01-10T09:58:20 & 2457032.924 & 0.70 & 0.045$\pm$0.006 & 0.042$\pm$0.006 \\ 
104$^-$ &\swift & 00032960112 (1.4 ks) & 2015-01-10T13:26:00 & 2457033.068 & 0.71 & 0.041$\pm$0.006 & 0.045$\pm$0.006 \\ 
105$^-$ &\swift & 00032960113 (1.5 ks) & 2015-01-10T16:28:13 & 2457033.195 & 0.71 & 0.035$\pm$0.005 & 0.044$\pm$0.006 \\ 
106$^-$ &\swift & 00032960114 (1.4 ks) & 2015-01-10T21:32:20 & 2457033.430 & 0.72 & 0.048$\pm$0.006 & 0.039$\pm$0.006 \\ 
107$^-$ &\swift & 00032960115 (1.5 ks) & 2015-01-11T00:38:17 & 2457033.566 & 0.72 & 0.031$\pm$0.005 & 0.036$\pm$0.005 \\ 
108     &\swift & 00032960116 (1.2 ks) & 2015-01-11T05:39:00 & 2457033.762 & 0.73 & 0.044$\pm$0.007 & 0.038$\pm$0.007 \\ 
109$^-$ &\swift & 00032960117 (1.4 ks) & 2015-01-11T09:57:39 & 2457033.923 & 0.73 & 0.046$\pm$0.006 & 0.039$\pm$0.006 \\ 
110$^-$ &\swift & 00032960118 (1.5 ks) & 2015-01-11T13:20:19 & 2457034.064 & 0.74 & 0.047$\pm$0.006 & 0.035$\pm$0.005 \\ 
111$^-$ &\swift & 00032960119 (1.5 ks) & 2015-01-11T16:33:19 & 2457034.198 & 0.74 & 0.040$\pm$0.006 & 0.041$\pm$0.006 \\ 
112$^-$ &\swift & 00032960120 (1.5 ks) & 2015-01-11T21:11:24 & 2457034.392 & 0.75 & 0.039$\pm$0.006 & 0.038$\pm$0.006 \\ 
113$^-$ &\swift & 00032960121 (1.4 ks) & 2015-01-12T00:36:17 & 2457034.563 & 0.75 & 0.038$\pm$0.006 & 0.037$\pm$0.006 \\ 
114     &\swift & 00032960122 (1.3 ks) & 2015-01-12T05:35:56 & 2457034.760 & 0.76 & 0.038$\pm$0.007 & 0.041$\pm$0.007 \\ 
115$^-$ &\swift & 00032960123 (1.3 ks) & 2015-01-12T10:09:17 & 2457034.931 & 0.76 & 0.047$\pm$0.006 & 0.039$\pm$0.006 \\ 
116$^-$ &\swift & 00032960124 (1.5 ks) & 2015-01-12T13:18:12 & 2457035.063 & 0.77 & 0.054$\pm$0.008 & 0.040$\pm$0.007 \\ 
117$^-$ &\swift & 00032960125 (1.5 ks) & 2015-01-12T16:29:13 & 2457035.196 & 0.77 & 0.062$\pm$0.007 & 0.059$\pm$0.007 \\ 
118$^-$ &\swift & 00032960126 (1.5 ks) & 2015-01-12T22:52:17 & 2457035.462 & 0.78 & 0.055$\pm$0.007 & 0.040$\pm$0.006 \\ 
119$^-$ &\swift & 00032960127 (1.5 ks) & 2015-01-13T00:33:12 & 2457035.561 & 0.78 & 0.048$\pm$0.007 & 0.038$\pm$0.006 \\ 
120$^-$ &\swift & 00032960128 (1.0 ks) & 2015-01-13T05:04:09 & 2457035.749 & 0.79 & 0.051$\pm$0.008 & 0.030$\pm$0.006 \\ 
\hline
\end{tabular}
\end{table*}
\setcounter{table}{0}
\begin{table*}
\centering
\footnotesize
\caption{Continued.}
\begin{tabular}{llcccccc}
\hline\hline
ID & Obs. & ObsID (exp. time) & Start Date & HJD & $\phi$ & \multicolumn{2}{c}{Count Rates (ct\,s$^{-1}$)}\\
   &      &                   &            & at mid-exposure &        & 0.4--2.0\,keV & 2.0--10.0\,keV \\
\hline
121$^-$ &\swift & 00032960129 (1.4 ks) & 2015-01-13T10:04:44 & 2457035.928 & 0.80 & 0.034$\pm$0.005 & 0.044$\pm$0.006 \\ 
122$^-$ &\swift & 00032960130 (1.4 ks) & 2015-01-13T13:17:59 & 2457036.062 & 0.80 & 0.047$\pm$0.007 & 0.036$\pm$0.006 \\ 
123$^-$ &\swift & 00032960131 (1.4 ks) & 2015-01-13T16:29:00 & 2457036.195 & 0.80 & 0.055$\pm$0.007 & 0.041$\pm$0.006 \\ 
124     &\swift & 00032960133 (0.6 ks) & 2015-01-14T02:13:31 & 2457036.596 & 0.82 & 0.067$\pm$0.012 & 0.034$\pm$0.008 \\ 
125$^-$ &\swift & 00032960134 (1.3 ks) & 2015-01-14T04:59:20 & 2457036.777 & 0.82 & 0.055$\pm$0.007 & 0.048$\pm$0.007 \\ 
126     &\swift & 00032960135 (0.3 ks) & 2015-01-14T10:03:00 & 2457036.927 & 0.83 & 0.033$\pm$0.012 & 0.036$\pm$0.012 \\ 
127$^-$ &\swift & 00032960136 (1.5 ks) & 2015-01-14T12:59:52 & 2457037.051 & 0.83 & 0.050$\pm$0.006 & 0.053$\pm$0.006 \\ 
128$^-$ &\swift & 00032960137 (1.3 ks) & 2015-01-14T16:28:00 & 2457037.194 & 0.84 & 0.052$\pm$0.007 & 0.042$\pm$0.006 \\ 
129$^-$ &\swift & 00032960138 (1.4 ks) & 2015-01-14T21:13:58 & 2457037.393 & 0.84 & 0.052$\pm$0.007 & 0.054$\pm$0.007 \\ 
130$^*$ &\swift & 00032960139 (1.5 ks) & 2015-01-15T00:28:24 & 2457037.557 & 0.85 & 0.068$\pm$0.008 & 0.070$\pm$0.008 \\ 
131$^-$ &\swift & 00032960141 (1.1 ks) & 2015-01-15T11:25:54 & 2457037.983 & 0.86 & 0.051$\pm$0.008 & 0.050$\pm$0.008 \\ 
132$^-$ &\swift & 00032960142 (1.2 ks) & 2015-01-15T14:35:51 & 2457038.116 & 0.87 & 0.068$\pm$0.008 & 0.056$\pm$0.007 \\ 
133$^-$ &\swift & 00032960143 (1.3 ks) & 2015-01-15T17:45:18 & 2457038.248 & 0.87 & 0.060$\pm$0.007 & 0.062$\pm$0.007 \\ 
134$^-$ &\swift & 00032960144 (1.0 ks) & 2015-01-15T21:11:38 & 2457038.389 & 0.87 & 0.074$\pm$0.009 & 0.055$\pm$0.008 \\ 
135$^-$ &\swift & 00032960145 (1.4 ks) & 2015-01-16T00:34:07 & 2457038.560 & 0.88 & 0.048$\pm$0.007 & 0.035$\pm$0.006 \\ 
136     &\swift & 00032960146 (0.6 ks) & 2015-01-16T04:59:32 & 2457038.753 & 0.89 & 0.040$\pm$0.011 & 0.034$\pm$0.010 \\ 
137$^-$ &\swift & 00032960147 (0.9 ks) & 2015-01-16T09:47:07 & 2457038.913 & 0.89 & 0.052$\pm$0.008 & 0.062$\pm$0.009 \\ 
138     &\swift & 00032960148 (0.4 ks) & 2015-01-16T13:10:18 & 2457039.051 & 0.89 & 0.065$\pm$0.018 & 0.074$\pm$0.018 \\ 
139     &\swift & 00032960149 (0.8 ks) & 2015-01-16T17:53:09 & 2457039.250 & 0.90 & 0.076$\pm$0.011 & 0.050$\pm$0.009 \\ 
140$^-$ &\swift & 00032960150 (1.1 ks) & 2015-01-16T21:08:36 & 2457039.388 & 0.91 & 0.072$\pm$0.009 & 0.074$\pm$0.009 \\ 
141$^-$ &\swift & 00032960151 (1.3 ks) & 2015-01-17T00:25:17 & 2457039.555 & 0.91 & 0.061$\pm$0.008 & 0.067$\pm$0.008 \\ 
142     &\swift & 00032960153 (1.0 ks) & 2015-01-17T09:46:15 & 2457039.914 & 0.92 & 0.073$\pm$0.013 & 0.084$\pm$0.014 \\ 
143$^-$ &\swift & 00032960154 (1.0 ks) & 2015-01-17T14:29:42 & 2457040.110 & 0.93 & 0.052$\pm$0.008 & 0.067$\pm$0.009 \\ 
144$^*$ &\swift & 00032960155 (1.4 ks) & 2015-01-17T17:42:38 & 2457040.247 & 0.93 & 0.069$\pm$0.007 & 0.065$\pm$0.007 \\ 
145     &\swift & 00032960156 (0.2 ks) & 2015-01-17T20:59:27 & 2457040.384 & 0.94 & 0.038$\pm$0.016 & 0.045$\pm$0.017 \\ 
146     &\swift & 00032960157 (0.6 ks) & 2015-01-18T00:19:56 & 2457040.519 & 0.94 & 0.056$\pm$0.016 & 0.071$\pm$0.017 \\ 
147     &\swift & 00032960158 (0.7 ks) & 2015-01-18T06:30:01 & 2457040.775 & 0.95 & 0.050$\pm$0.009 & 0.065$\pm$0.010 \\ 
148$^-$ &\swift & 00032960159 (1.4 ks) & 2015-01-18T11:25:39 & 2457040.985 & 0.96 & 0.045$\pm$0.006 & 0.060$\pm$0.007 \\ 
149$^-$ &\swift & 00032960160 (1.3 ks) & 2015-01-18T14:37:03 & 2457041.117 & 0.96 & 0.053$\pm$0.007 & 0.049$\pm$0.007 \\ 
150$^-$ &\swift & 00032960161 (1.4 ks) & 2015-01-18T17:47:20 & 2457041.250 & 0.96 & 0.045$\pm$0.006 & 0.057$\pm$0.007 \\ 
151     &\swift & 00032960162 (1.5 ks) & 2015-01-18T22:37:17 & 2457041.452 & 0.97 & 0.017$\pm$0.004 & 0.035$\pm$0.006 \\ 
152     &\swift & 00032960163 (1.1 ks) & 2015-01-19T01:56:59 & 2457041.619 & 0.98 & 0.016$\pm$0.004 & 0.021$\pm$0.005 \\ 
153     &\swift & 00032960164 (1.2 ks) & 2015-01-19T06:23:54 & 2457041.774 & 0.98 & 0.014$\pm$0.005 & 0.027$\pm$0.007 \\ 
154     &\swift & 00032960165 (1.4 ks) & 2015-01-19T09:47:11 & 2457041.916 & 0.99 & 0.007$\pm$0.003 & 0.017$\pm$0.004 \\ 
155     &\swift & 00032960166 (1.5 ks) & 2015-01-19T14:23:36 & 2457042.109 & 0.99 & 0.004$\pm$0.003 & 0.014$\pm$0.005 \\ 
156     &\swift & 00032960167 (1.5 ks) & 2015-01-19T17:44:49 & 2457042.249 & 1.00 & 0.008$\pm$0.003 & 0.008$\pm$0.003 \\ 
157     &\swift & 00032960168 (1.4 ks) & 2015-01-19T22:35:30 & 2457042.450 & 0.00 & 0.009$\pm$0.003 & 0.007$\pm$0.003 \\ 
158     &\swift & 00032960169 (1.2 ks) & 2015-01-20T00:14:56 & 2457042.549 & 0.01 & 0.007$\pm$0.003 & 0.013$\pm$0.004 \\ 
159     &\swift & 00032960170 (1.1 ks) & 2015-01-20T06:29:08 & 2457042.777 & 0.01 & 0.013$\pm$0.006 & 0.002$\pm$0.003 \\ 
160     &\swift & 00032960171 (0.3 ks) & 2015-01-20T11:27:19 & 2457042.979 & 0.02 & 0.008$\pm$0.007 & 0.028$\pm$0.011 \\ 
161     &\swift & 00032960172 (0.6 ks) & 2015-01-20T14:36:15 & 2457043.113 & 0.02 & 0.019$\pm$0.008 & 0.027$\pm$0.009 \\ 
162     &\swift & 00032960173 (0.3 ks) & 2015-01-20T17:50:48 & 2457043.246 & 0.03 & 0.020$\pm$0.009 & 0.012$\pm$0.007 \\ 
163     &\swift & 00032960174 (0.8 ks) & 2015-01-20T20:57:22 & 2457043.378 & 0.03 & 0.025$\pm$0.006 & 0.015$\pm$0.005 \\ 
164     &\swift & 00032960175 (0.9 ks) & 2015-01-21T00:12:08 & 2457043.516 & 0.04 & 0.018$\pm$0.005 & 0.026$\pm$0.006 \\ 
165     &\swift & 00032960176 (1.5 ks) & 2015-01-21T06:19:02 & 2457043.772 & 0.04 & 0.012$\pm$0.003 & 0.023$\pm$0.004 \\ 
166     &\swift & 00032960177 (1.3 ks) & 2015-01-21T11:06:02 & 2457043.971 & 0.05 & 0.023$\pm$0.005 & 0.024$\pm$0.005 \\ 
167     &\swift & 00032960178 (1.3 ks) & 2015-01-21T14:17:24 & 2457044.104 & 0.05 & 0.023$\pm$0.005 & 0.027$\pm$0.005 \\ 
168     &\swift & 00032960179 (1.3 ks) & 2015-01-21T17:40:24 & 2457044.244 & 0.06 & 0.029$\pm$0.005 & 0.026$\pm$0.005 \\ 
169     &\swift & 00032960180 (1.2 ks) & 2015-01-21T20:54:42 & 2457044.379 & 0.06 & 0.026$\pm$0.005 & 0.022$\pm$0.005 \\ 
170     &\swift & 00032960181 (0.2 ks) & 2015-01-22T03:04:57 & 2457044.630 & 0.07 & 0.025$\pm$0.011 & 0.045$\pm$0.015 \\ 
171     &\swift & 00032960182 (0.4 ks) & 2015-01-22T06:24:55 & 2457044.770 & 0.08 & 0.030$\pm$0.010 & 0.018$\pm$0.007 \\ 
172     &\swift & 00032960183 (0.9 ks) & 2015-01-22T11:04:49 & 2457044.968 & 0.08 & 0.019$\pm$0.006 & 0.030$\pm$0.008 \\ 
173     &\swift & 00032960184 (0.6 ks) & 2015-01-22T12:40:53 & 2457045.033 & 0.08 & 0.042$\pm$0.009 & 0.030$\pm$0.007 \\ 
174     &\swift & 00032960185 (1.5 ks) & 2015-01-22T17:33:13 & 2457045.241 & 0.09 & 0.036$\pm$0.006 & 0.031$\pm$0.006 \\ 
175     &\swift & 00032960186 (1.2 ks) & 2015-01-22T20:48:06 & 2457045.374 & 0.09 & 0.037$\pm$0.006 & 0.032$\pm$0.006 \\ 
176     &\swift & 00032960187 (0.1 ks) & 2015-01-23T03:04:24 & 2457045.630 & 0.10 & 0.025$\pm$0.015 & 0.058$\pm$0.022 \\ 
177     &\swift & 00032960188 (0.3 ks) & 2015-01-23T06:14:14 & 2457045.762 & 0.11 & 0.044$\pm$0.014 & 0.022$\pm$0.010 \\ 
178$^-$ &\swift & 00032960190 (2.0 ks) & 2015-01-23T12:56:49 & 2457046.104 & 0.12 & 0.047$\pm$0.006 & 0.036$\pm$0.005 \\ 
179     &\swift & 00032960194 (0.5 ks) & 2015-01-24T06:14:49 & 2457046.764 & 0.14 & 0.037$\pm$0.010 & 0.035$\pm$0.010 \\ 
180     &\swift & 00032960195 (0.6 ks) & 2015-01-24T08:01:44 & 2457046.873 & 0.14 & 0.026$\pm$0.008 & 0.035$\pm$0.009 \\ 
\hline
\end{tabular}
\end{table*}
\setcounter{table}{0}
\begin{table*}
\centering
\footnotesize
\caption{Continued.}
\begin{tabular}{llcccccc}
\hline\hline
ID & Obs. & ObsID (exp. time) & Start Date & HJD & $\phi$ & \multicolumn{2}{c}{Count Rates (ct\,s$^{-1}$)}\\
   &      &                   &            & at mid-exposure &        & 0.4--2.0\,keV & 2.0--10.0\,keV \\
\hline
181$^-$ &\swift & 00032960196 (1.5 ks) & 2015-01-24T12:37:44 & 2457047.064 & 0.15 & 0.037$\pm$0.006 & 0.033$\pm$0.005 \\ 
182     &\swift & 00032960197 (0.5 ks) & 2015-01-24T19:02:22 & 2457047.297 & 0.15 & 0.045$\pm$0.010 & 0.039$\pm$0.009 \\ 
183     &\swift & 00032960198 (0.4 ks) & 2015-01-24T22:37:13 & 2457047.446 & 0.16 & 0.046$\pm$0.012 & 0.049$\pm$0.012 \\ 
184     &\swift & 00032960199 (0.4 ks) & 2015-01-25T00:03:17 & 2457047.566 & 0.16 & 0.028$\pm$0.010 & 0.028$\pm$0.010 \\ 
185     &\swift & 00032960201 (1.3 ks) & 2015-01-25T09:24:42 & 2457047.901 & 0.17 & 0.037$\pm$0.006 & 0.030$\pm$0.005 \\ 
186     &\swift & 00032960202 (0.7 ks) & 2015-01-25T12:38:30 & 2457048.032 & 0.18 & 0.045$\pm$0.009 & 0.036$\pm$0.008 \\ 
187     &\swift & 00032960203 (0.6 ks) & 2015-01-25T17:49:26 & 2457048.270 & 0.19 & 0.039$\pm$0.009 & 0.035$\pm$0.008 \\ 
188$^-$ &\swift & 00032960207 (1.4 ks) & 2015-01-26T09:26:16 & 2457048.902 & 0.21 & 0.039$\pm$0.006 & 0.031$\pm$0.005 \\ 
189     &\swift & 00032960208 (0.7 ks) & 2015-01-26T14:10:16 & 2457049.126 & 0.21 & 0.033$\pm$0.008 & 0.040$\pm$0.008 \\ 
190     &\swift & 00032960209 (0.1 ks) & 2015-01-26T17:21:14 & 2457049.225 & 0.22 & 0.086$\pm$0.034 & 0.049$\pm$0.026 \\ 
191     &\swift & 00032960211 (1.0 ks) & 2015-01-27T00:03:17 & 2457049.537 & 0.23 & 0.037$\pm$0.007 & 0.030$\pm$0.006 \\ 
192     &\swift & 00032960212 (0.8 ks) & 2015-01-27T07:44:14 & 2457049.828 & 0.23 & 0.050$\pm$0.010 & 0.030$\pm$0.008 \\ 
193     &\swift & 00032960213 (1.3 ks) & 2015-01-27T11:02:03 & 2457049.968 & 0.24 & 0.030$\pm$0.006 & 0.033$\pm$0.007 \\ 
194     &\swift & 00032960214 (1.0 ks) & 2015-01-27T12:34:03 & 2457050.061 & 0.24 & 0.038$\pm$0.007 & 0.028$\pm$0.006 \\ 
195$^-$ &\swift & 00032960215 (1.5 ks) & 2015-01-27T17:20:13 & 2457050.232 & 0.25 & 0.038$\pm$0.006 & 0.032$\pm$0.005 \\ 
196     &\swift & 00032960216 (0.4 ks) & 2015-01-27T23:53:12 & 2457050.499 & 0.26 & 0.029$\pm$0.010 & 0.033$\pm$0.011 \\ 
197     &\swift & 00032960217 (0.7 ks) & 2015-01-28T01:27:15 & 2457050.566 & 0.26 & 0.050$\pm$0.009 & 0.029$\pm$0.007 \\ 
198     &\swift & 00032960218 (1.2 ks) & 2015-01-28T07:51:02 & 2457050.835 & 0.27 & 0.037$\pm$0.006 & 0.028$\pm$0.005 \\ 
199     &\swift & 00032960219 (1.5 ks) & 2015-01-28T09:20:22 & 2457050.899 & 0.27 & 0.037$\pm$0.006 & 0.026$\pm$0.005 \\ 
200     &\swift & 00032960220 (1.5 ks) & 2015-01-28T12:33:29 & 2457051.033 & 0.27 & 0.052$\pm$0.008 & 0.035$\pm$0.007 \\ 
201$^-$ &\swift & 00032960221 (1.4 ks) & 2015-01-28T17:21:29 & 2457051.233 & 0.28 & 0.041$\pm$0.006 & 0.039$\pm$0.006 \\ 
202     &\swift & 00032960222 (0.6 ks) & 2015-01-28T20:47:19 & 2457051.371 & 0.28 & 0.061$\pm$0.013 & 0.025$\pm$0.009 \\ 
203$^*$ &{\it Chandra}& 9113 (4.7ks)   & 2008-04-27T00:56:56 & 2454583.580 & 0.39 & 0.116$\pm$0.005 & 0.095$\pm$0.005 \\
\hline
\end{tabular}
\tablefoot{
$^*$ observation leading to a good spectrum; $^-$ observation leading to a rough spectrum with about 80--150 raw counts for the source; both $^*$ and $^-$ cases were used for spectral fitting (see Sect.~4 and 
Table~\ref{spectralfits}).}
\end{table*}
\end{appendix}

\begin{thebibliography}{00}
%
\bibitem[Ackermann et al.(2011)]{acke11} Ackermann, M., Ajello, M., Baldini, L., et al.\ 2011, \apj, 726, 35 
\bibitem[Ascenso et al.(2007)]{asce07} Ascenso, J., Alves, J., Beletsky, Y., \& Lago, M.T.V.T.\ 2007, \aap, 466, 137 
\bibitem[Asplund et al.(2009)]{aspl09} 
Asplund, M., Grevesse, N., Sauval, A.J., \& Scott, P.\ 2009, 
\araa, 47, 481 
\bibitem[Baluci\'nska-Church \& McCammon(1992)]{bal92} Baluci\'nska-Church, M., \& McCammon, D.\ 1992, \apj, 400, 699 
\bibitem[Belloni \& Mereghetti(1994)]{beme94}
Belloni, T., \& Mereghetti, S. 1994, \aap, 286, 935 
\bibitem[Benaglia et al.(2005)]{bena05}
Benaglia, P., Romero, G.E., Koribalski, B., \& Pollock, A.M.T.
2005, \aap, 440, 743
\bibitem[Bergh\"ofer et al.(1997)]{berg97}
Bergh\"ofer, T.W., Schmitt, J.H.M.M., Danner, R., \& Cassinelli, J.P.
1997, \aap, 322, 167 
\bibitem[Bohlin et al.(1978)]{bohl78}
Bohlin, R.C., Savage, B.D., \& Drake, J.F. 1978, \apj, 224, 132 
\bibitem[Caraveo(1983)]{cara83}
Caraveo, P.A. 1983, \ssr, 36, 207
\bibitem[Caraveo et al.(1989)]{cara89}
Caraveo, P.A., Bignami, G.F., \& Goldwurm, A. 1989, \apj, 338, 338
\bibitem[Cazorla et al.(2014)]{cazo14}
Cazorla, C., Naz\'e, Y., \& Rauw, G. 2014, \aap, 561, A92
\bibitem[Corcoran et al.(2011)]{corc11}
Corcoran, M.F., Pollock, A.M.T., Hamaguchi, K., \& Russell, C.
2011, arXiv:1101.1422
\bibitem[Crowther et al.(1995)]{crow95}
Crowther, P.A., Hillier, D.J., \& Smith, L.J. 1995, \aap, 293, 403
\bibitem[De Becker \& Raucq(2013)]{dera13}
De Becker, M., \& Raucq, F. 2013, \aap, 558, A28 
\bibitem[Dieters et al.(1990)]{diet90}
Dieters, S.W., Hill, K.M., \& Watson, R.D. 1990, 
Inf. Bull. of Var. Stars, 3500, 1
\bibitem[Dorman \& Arnaud(2001)]{dor01} Dorman, B., \& Arnaud, K.~A.\ 2001, Astronomical Data Analysis Software and Systems X, ASPC vol.238, 415 
\bibitem[Feldmeier et al.(1997a)]{feld97a}
Feldmeier, A., Kudritzki, R.P., Palsa, R., et al.\ 1997a, \aap, 320, 899
\bibitem[Feldmeier et al.(1997b)]{feld97b}
Feldmeier, A., Puls, J., \& Pauldrach, A.W.A. 1997b, \aap, 322, 878
\bibitem[Fich et al.(1989)]{fich89}
Fich, M., Blitz, L., \& Stark, A.A. 1989, \apj, 342, 272
\bibitem[Fukui et al.(2009)]{fuku09} 
Fukui, Y., Furukawa, N., Dame, T.~M., et al.\ 2009, \pasj, 61, L23 
\bibitem[Gamen et al.(2006)]{gago06}
Gamen, R., Gosset, E., Morrell, N., et al.\ 2006, \aap, 460, 777
\bibitem[Gamen et al.(2009)]{gafe09}
Gamen, R., Fern{\'a}ndez-Laj{\'u}s, E., Niemela, V.S., \& Barb{\'a}, R.H. 
2009, \aap, 506, 1269 
\bibitem[Gayley et al.(1997)]{gayl97}
Gayley, K.G., Owocki, S.P., \& Cranmer, S.R. 1997, \apj, 475, 786
\bibitem[Golwurm et al.(1987)]{gold87}
Goldwurm, A., Caraveo, P.A., \& Bignami, G.F. 1987, \apj, 322, 349
\bibitem[Gosset(2007)]{goss07}
Gosset, E. 2007, Habilitation Thesis, Li\`ege University, Belgium
\bibitem[Gosset et al.(2005)]{goss05}
Gosset, E., Naz\'e, Y., Claeskens, J.F., et al.\ 2005, \aap,
429, 685
\bibitem[Gosset et al.(2009)]{goss09}
Gosset, E., Naz\'e, Y., Sana, H., et al.\ 2009, \aap, 508, 805
\bibitem[Gr\"afener \& Hamann(2008)]{graf08}
Gr\"afener, G., \& Hamann, W.R. 2008, \aap, 482, 945
\bibitem[G\"udel \& Naz\'e(2010)]{guna10}
G\"udel, M., \& Naz\'e, Y. 2010, \ssr, 157, 211
\bibitem[Gudennavar et al.(2012)]{gud12} 
Gudennavar, S.~B., Bubbly, S.~G., Preethi, K., 
\& Murthy, J.\ 2012, \apjs, 199, 8 
\bibitem[Hamann et al.(2006)]{hama06}
Hamann, W.R., Gr\"afener, G., \& Liermann, A. 2006, \aap, 457, 1015
\bibitem[Harnden et al.(1979)]{harn79}
Harnden, F.R.Jr., Branduardi, G., Elvis, M., et al.\ 1979, \apj, 234, L51
\bibitem[Hertz \& Grindlay(1984)]{hegr84}
Hertz, P., \& Grindlay, J.E. 1984, \apj, 278, 137 
\bibitem[Hillier \& Miller(1998)]{hil98} Hillier, D.~J., \& Miller, D.~L.\ 1998, \apj, 496, 407 
\bibitem[Kaufer et al.(1999)]{kauf99}
Kaufer, A., Stahl, O., Tubbesing, S., et al.\ 1999, ESO Msngr, 95, 8 
\bibitem[Lindgren(1976)]{lind76} Lindgren, B.W.\ 1976, Statistical theory - third edition, McMillan Pub. (New York)
\bibitem[Lomax et al.(2015)]{loma15} 
Lomax, J.~R., Naz{\'e}, Y., Hoffman, J.~L., et al.\ 2015, \aap, 573, A43 
\bibitem[Lucy(1982)]{lucy82}
Lucy, L.B. 1982, \apj, 255, 286
\bibitem[Lucy \& White(1980)]{luwh80}
Lucy, L.B., \& White, R.L. 1980, \apj, 241, 300 
\bibitem[Mahy et al.(2012)]{mago12}
Mahy, L., Gosset, E., Sana, H., et al.\ 2012, \aap, 540, A97
\bibitem[Martins et al.(2005)]{mart05}
Martins, F., Schaerer, D., \& Hillier, D.J. 2005, \aap, 436, 1049 
\bibitem[Mereghetti \& Belloni(1994)]{mebe94}
Mereghetti, S., \& Belloni, T. 1994, \memsai, 65, 369
\bibitem[Mereghetti et al.(1994)]{mere94}
Mereghetti, S., Belloni, T., Shara, M., \& Drissen, L.
1994, \apj, 424, 943 
\bibitem[Mereghetti et al.(1995)]{mere95}
Mereghetti, S., Belloni, T., Haberl, F., \& Voges, W. 1995,
in Wolf-Rayet stars: Binaries, Colliding Winds, Evolution,
eds. K.A. van der Hucht \& P.M. Williams,
proceedings of IAU Symposium no.163, p.481
\bibitem[Moffat \& Corcoran(2009)]{moco09}
Moffat, A.F.J., \& Corcoran, M.F. 2009, \apj, 707, 693
\bibitem[Muijres et al.(2012)]{muij12}
Muijres, L.E., Vink, J.S., de Koter, A., et al.\ 2012, \aap, 537, A37 
\bibitem[Naz{\'e}(2009)]{naze09}
Naz{\'e}, Y. 2009, \aap, 506, 1055
\bibitem[Naz{\'e} et al.(2002)]{naze02}
Naz{\'e}, Y., Hartwell, J.M., Stevens, I.R., et al.\ 2002,
\apj, 580, 225
\bibitem[Naz{\'e} et al.(2008)]{naze08}
Naz{\'e}, Y., Rauw, G., \& Manfroid, J.\ 2008, \aap, 483, 171
\bibitem[Naz{\'e} et al.(2013)]{naze13} 
Naz{\'e}, Y., Oskinova, L.M., \& Gosset, E.\ 2013, \apj, 763, 143 
\bibitem[Niemela et al.(2006)]{niem06}
Niemela, V.S., Gamen, R.C., Solivella, G.R. et al.\ 2006, 
Rev.\ Mex.\ A\&A (Conf.\ Ser.), 26, 39 
\bibitem[Niemela et al.(2008)]{niem08} 
Niemela, V.S., Gamen, R.C., Barb{\'a}, R.H., et al.\ 2008, \mnras, 389, 1447 
\bibitem[Oskinova et al.(2003)]{oski03}
Oskinova, L.M., Ignace, R., Hamann, W.R., et al.\ 2003, \aap, 402, 755
\bibitem[Pallavicini et al.(1981)]{pall81}
Pallavicini, R., Golub, L., Rosner, R., et al.\ 1981, \apj, 248, 279
\bibitem[Pandey et al.(2014)]{pand14}
Pandey, J.C., Pandey, S.B., \& Karnakar, S. 2014, \apj, 788, 84
\bibitem[Parkin \& Gosset(2011)]{park11} 
Parkin, E.R., \& Gosset, E.\ 2011, \aap, 530, A119 
\bibitem[Pittard \& Parkin(2010)]{pipa10}
Pittard, J.M., \& Parkin, E.R. 2010, \mnras, 403, 1657
\bibitem[Rauw et al.(1996)]{rauw96}
Rauw, G., Vreux, J.M., Gosset, E., et al.\ 1996, \aap, 306, 771
\bibitem[Rauw et al.(2005)]{rauw05}
Rauw, G., Crowther, P.A., De Becker, M., et al.\ 2005,
\aap, 432, 985
\bibitem[Rauw et al.(2011)]{rauw11} 
Rauw, G., Sana, H., \& Naz{\'e}, Y.\ 2011, \aap, 535, A40
\bibitem[Rauw et al.(2016)]{ramo16}
Rauw, G., Mossoux, E., \& Naz\'e, Y. 2016, \na, 43, 70
\bibitem[Reig(1999)]{reig99}
Reig, P. 1999, \aap, 345, 576
\bibitem[Roberts et al.(2001)]{robe01}
Roberts, M.S.E., Romani, R.W., \& Kawai, N. 2001, \apjs, 133, 451 
\bibitem[Roman-Lopes, Barba, \& Morrell(2011)]{roma11}
Roman-Lopes, A., Barba, R.H., \& Morrell, N.I. 2011, \mnras, 416, 501
\bibitem[Rosslowe \& Crowther(2015)]{rocr15}
Rosslowe, C.K., \& Crowther, P.A. 2015, \mnras, 447, 2322
(Erratum: 449, 2436) 
\bibitem[Sana et al.(2006)]{sana06}
Sana, H., Rauw, G., Naz\'e, Y., et al.\
2006, \mnras, 372, 661
\bibitem[Seward \& Chlebowski(1982)]{sech82}
Seward, F.D., \& Chlebowski, T. 1982, \apj, 256, 530 
\bibitem[Seward et al.(1979)]{sewa79}
Seward, F.D., Forman, W.R., Giacconi, R., et al.\ 1979, \apj, 234, L55 
\bibitem[Skinner et al.(2010)]{skin10}
Skinner, S.L., Zhekov, S.A., G\"udel, M., et al.\ 2010, \aj, 139, 825
\bibitem[Stevens et al.(1992)]{stev92}
Stevens, I.R., Blondin, J.M., \& Pollock, A.M.T. 1992, \apj, 386, 265  
\bibitem[Sugawara et al.(2015)]{suga15}
Sugawara, Y., Tsuboi, Y., Maeda, Y., Pollock, A.M.T., 
\& Williams, P.M. 2015, in International Workshop on Wolf-Rayet Stars,
eds. W.R.\ Hamann, A.\ Sander, \& H.\ Todt, Potsdam University
Press 2016, p.366
\bibitem[The(1966)]{thee66}
The, P.S. 1966, Contr. from the Bosscha Observ., Lembang, 35
\bibitem[Tramper et al.(2016)]{tram16}
Tramper, F., Sana, H., Fitzsimons, N.E., et al.\ 2016, \mnras, 455, 1275
\bibitem[Vargas {\'A}lvarez et al.(2013)]{varg13} 
Vargas {\'A}lvarez, C.~A., Kobulnicky, H.~A., Bradley, D.~R., 
et al.\ 2013, \aj, 145, 125 
\bibitem[Wackerling(1970)]{wack70}
Wackerling, L.R. 1970, \memras, 73, 153
\bibitem[Wallerstein, Sandstrom, \& Gredel(2007)]{wasa07}
Wallerstein, G., Sandstrom, K., \& Gredel, R. 2007, \pasp, 119, 1268
\bibitem[Yan et al.(1998)]{bal98} Yan, M., Sadeghpour, H.~R., 
\& Dalgarno, A.\ 1998, \apj, 496, 1044 
\bibitem[Zacharias et al.(2013)]{zach13}
Zacharias, N., Finch, C.T., Girard, T.M., et al.\ 2013, \aj, 145, 44
\bibitem[Zhekov(2012)]{zhek12}
Zhekov, S.A. 2012, \mnras, 422, 1332 
\end{thebibliography}
\end{document}